\documentclass[10pt, twocolumn]{IEEEtran}
\usepackage{amsmath}
\usepackage{relsize}
\usepackage{amssymb}
\usepackage{amsthm}
\usepackage{cases}
\usepackage{setspace}
\usepackage[font={footnotesize}]{caption}
\usepackage[export]{adjustbox}
\usepackage{graphicx}
\usepackage{epstopdf}
\usepackage{color}
\usepackage{hyperref}
\hypersetup{colorlinks = true, urlcolor = cyan}

\usepackage{url}

\newcommand\blfootnote[1]{%
  \begingroup
  \renewcommand\thefootnote{}\footnote{#1}%
  \addtocounter{footnote}{-1}%
  \endgroup
}

\newtheorem{theorem}{\bf Theorem}
\newtheorem{lemma}{\bf Lemma}

\newtheorem{problem}{\bf Problem}

\newtheorem{definition}{\bf Definition}
\newtheorem{corollary}{\bf Corollary}

\newenvironment{proofof}[1]{{\bf Proof of #1:}}{\hfill \IEEEQEDopen}
\newenvironment{proofsketch}{{\sc{Proof Outline:}}}{\hfill \IEEEQEDopen \vspace{0.1cm}}

\newcommand{\rate}{\ensuremath{R}}
\title{{\huge On the Total Power Capacity of Regular-LDPC Codes with Iterative Message-Passing Decoders}}
\date{}
\author{Karthik $\text{Ganesan}^{\dagger}$, Pulkit $\text{Grover}^{\ddagger}$, Jan $\text{Rabaey}^{\S}$, and Andrea $\text{Goldsmith}^{\dagger}$
\thanks{$\dagger$ Electrical Engineering, Stanford University. 
$\ddagger$ Electrical and Computer Engineering, Carnegie Mellon University.  
$\S$ Electrical Engineering and Computer Science, University of California at Berkeley. (Email correspondence should be addressed to karthik3@stanford.edu.)}}
\begin{document}
\maketitle
\thispagestyle{empty}

\allowdisplaybreaks

\vspace{-1.5cm}
\begin{abstract}
Motivated by recently derived fundamental limits on total (transmit + decoding) power for coded communication with VLSI decoders, this paper investigates the scaling behavior of the minimum total power needed to communicate over AWGN channels as the target bit-error-probability tends to zero. We focus on regular-LDPC codes and iterative message-passing decoders. We analyze scaling behavior under two VLSI complexity models of decoding. One model abstracts power consumed in processing elements (``node model''), and another abstracts power consumed in wires which connect the processing elements (``wire model''). We prove that a coding strategy using regular-LDPC codes with Gallager-B decoding achieves order-optimal scaling of total power under the node model. However, we also prove that regular-LDPC codes and iterative message-passing decoders cannot meet existing fundamental limits on total power under the wire model. Further, if the transmit energy-per-bit is bounded, total power grows at a rate that is \emph{worse} than uncoded transmission. Complementing our theoretical results, we develop detailed physical models of decoding implementations using post-layout circuit simulations. Our theoretical and numerical results show that approaching fundamental limits on total power requires increasing the complexity of both the code design and the corresponding decoding algorithm as communication distance is increased or error-probability is lowered.
\end{abstract} 	

	\noindent
	
\blfootnote{Early results related to this paper were presented at the $2012$ Allerton Conference~\cite{Allerton12Paper} and IEEE Globecom $2012$~\cite{Globecom12Paper}.}

\vspace{-0.6cm}
\section{Introduction}
\label{sec:intro}	
	Intuitively, the concept of Shannon capacity captures how much 
	information can be communicated across a channel under specified resource constraints. 
	While the problem of approaching Shannon capacity under solely \emph{transmit power} 
	constraints is well understood, modern communication often takes place at 
	transmitter-receiver distances that are very short (e.g., on-chip 
	communication~\cite{stojanovic}, short distance wired communication~\cite{zhengyaJournal}, 
	and extremely-high-frequency short-range wireless communication~\cite{wigig}). 
	Empirically, it has been observed that at such short distances, the power required for 
	processing a signal at the transmitter/receiver circuitry can dominate 
	the power required for transmission, sometimes by orders of magnitude~\cite{zhengyaJournal,JSAC11Paper,Marcu}. 
	For instance, the power consumed in the decoding circuitry of multi-gigabit-per-second 
	communication systems can be hundreds of milliwatts or more (e.g.,~\cite{zhengyaJournal,ldpcpowerreduction}), 
	while the transmit power required is only tens of milliwatts~\cite{Marcu}. 
	Thus, transmit power constraints do not abstract the relevant power consumed in many modern systems. 

	Shannon capacity, complemented by modern coding-theoretic constructions~\cite{ModernCodingTheory}, 
	has provided a framework that is provably good for minimizing transmit power 
	(e.g.,~in power-constrained AWGN channels). In this work, we focus on a capacity question 
	that is motivated by \emph{total power}: at what maximum rate can one communicate across a 
	channel for a given total power, and a specified error-probability? Alternatively, 
	given a target communication rate and error-probability, what is the minimum required 
	total power? The first simplifying perspective to this problem was adopted in~\cite{goldsmithbahai,massaad}, where 
	all of the processing power components at the transmitter and the receiver were 
	lumped together. However, processing power is influenced heavily by the 
	specific modulation choice, coding strategy, equalization strategy, etc.~\cite{zhengyaJournal,JSAC11Paper}. 
	Even for a fixed communication strategy, processing power depends strongly on
	the implementation technology (e.g., $45$ nm CMOS) and the choice of circuit architecture. 

	Using theoretical models of VLSI implementations~\cite{thompsonthesis}, recent literature has 
	explored fundamental scaling limits~\cite{JSAC11Paper,ISIT12Paper,CISS11Paper,BlakeKschischang1} 
	on the transmit + decoding power consumed by error-correcting codes. These works abstract 
	energy consumed in processing nodes~\cite{JSAC11Paper} and wires~\cite{CISS11Paper,ISIT12Paper,BlakeKschischang1} 
	in the VLSI decoders, and show that there is a fundamental tradeoff between transmit and decoding power. 
	
\begin{figure}[htbp]
   	\centering
   	\includegraphics[width=0.9\columnwidth]{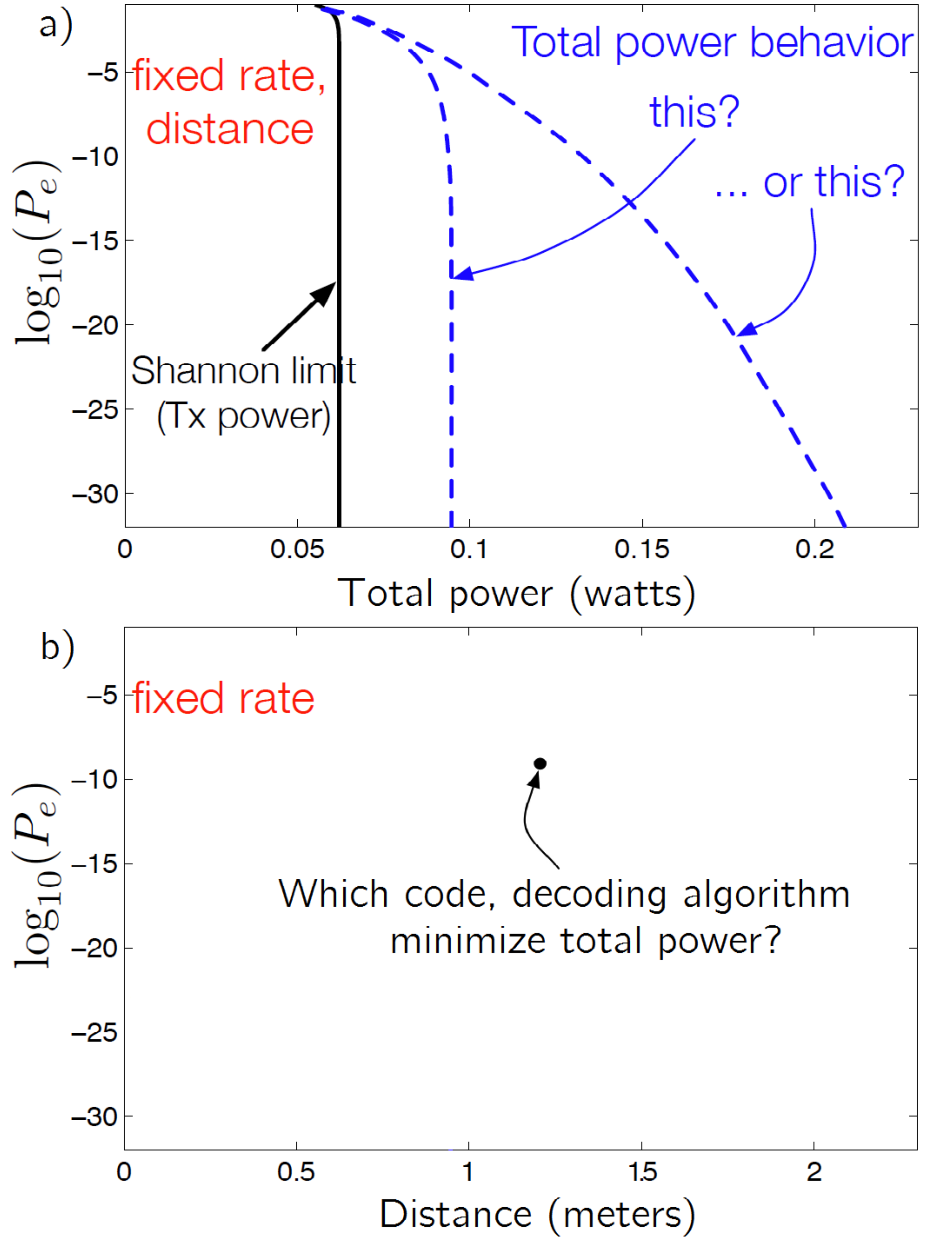}
   	\caption{a). The question explored in Sections~\ref{sec:asymptotic}-\ref{sec:wireanalysis}: 
	How fast does total power diverge to $\infty$ as bit-error probability $P_{e} \to 0$ for regular-LDPC
	codes and iterative message-passing decoding algorithms? b). The question explored in 
	Section~\ref{subsec:optimizationexample}: what is the most power-efficient pairing of a code and 
	decoding algorithm for a given distance and bit-error probability?}
   	\label{fig:question}
	\end{figure}

	In this work, we examine the achievability side of the question (see Fig.~\ref{fig:question}): 
	what is the total power that known code families and decoding algorithms can achieve? To address this 
	question, we first provide asymptotic bounds (Sections~\ref{sec:nodemodelanalysis}-\ref{sec:wireanalysis}) 
	on required decoding power. To do so, we restrict our analysis to binary regular-LDPC codes and iterative message-passing 
	decoding algorithms. Our code-family choice is motivated by both the order-optimality of regular-LDPC codes in some 
	theoretical models of circuit power~\cite{JSAC11Paper}, and their practical utility in both short~\cite{ethernetstandard} 
	and long~\cite{LDPCdeepspace} distance settings. Recent work of Blake and Kschischang~\cite{BlakeKschischang2} 
	also studied the energy complexity of LDPC decoding circuits, and an important connection to 
	this paper is highlighted in Section~\ref{sec:conclusion}.
	
	Within these restrictions we provide the following insights:
	\begin{enumerate}
	\item Wiring power, which explicitly brings out physical constraints 
	in a digital system~\cite{JanBook}, costs more in the order sense than the 
	power consumed in processing nodes. Thus, the commonly used metric for 
	decoding complexity --- number of operations --- underestimates circuit energy costs.	
	\item Shannon capacity is the maximal rate one can communicate at 
	with arbitrary reliability while the transmit power is held \emph{fixed}. 
	However, when total power minimization is the goal, keeping transmit power fixed while
	bit-error probability approaches zero can lead to highly suboptimal decoding power. 
	For instance, we prove that (Theorems~\ref{thm:anyLDPClower},~\ref{thm:gallageratotalpower},~\ref{thm:gallagebatotalpowerlower}) 
	at sufficiently low bit-error probability, it is more total power efficient to use \emph{uncoded transmission} 
	than regular-LDPC codes with iterative message-passing decoding, if using fixed transmit power. 
	However, if transmit power is allowed to diverge to infinity, we show that regular-LDPC codes 
	can outperform uncoded transmission in this total power sense.
	\item We prove (Corollary~\ref{cor:galbnode}) that a strategy using regular-LDPC codes 
	and the Gallager-B decoding achieves order-optimal scaling of total power when processing 
	power is dominated by nodes as opposed to wires (see Section~\ref{subsec:comparefundamental}).
	\item However, we also prove a lower bound (Theorem~\ref{thm:anyLDPClower}) that holds for all 
	regular-LDPC codes with iterative message-passing decoders for the case where processing power is 
	dominated by wires, and we show that a large gap exists between this lower bound and existing fundamental 
	limits (see Section~\ref{subsubsec:comparisonfundamentalwires}).
	\end{enumerate}
	
	To obtain insights on how an engineer might choose a power-efficient code for 
	a given system, we then develop empirical models of decoding power consumption of 
	$1$-bit and $2$-bit message-passing algorithms for regular-LDPC codes (Section~\ref{subsec:optimizationexample}). 
	These models are constructed using post-layout circuit simulations of power consumption 
	for check-node and variable-node sub-circuits, and generalizing the remaining 
	components of power to structurally similar codes. 

	Shannon-theoretic analysis yields transmit-power-centric results, 
	which are plotted as ``waterfall'' curves (with corresponding ``error-floors'') 
	demonstrating how close the code performs to the Shannon limit. 
	There, the channel path-loss can usually be ignored because it is merely a 
	scaling factor for the term to be optimized (namely the transmit power), 
	thereby not affecting the optimizing code. Since we are 
	interested in \emph{total} power, the path-loss impacts the code choice. 
	For simplicity of understanding, path-loss is translated into a more 
	relatable metric --- communication distance --- using a simple model 
	for path-loss. The resulting question is illustrated in Fig.~\ref{fig:question}(b): 
	At a given data-rate, what code and corresponding decoding algorithm minimize 
	the transmit + decoding power for a given transmit distance and bit-error probability?

	In Section~\ref{subsec:optimizationexample}, we present optimization 
	results for this question in a $60$ GHz communication setting using our models. 
	This particular setting is chosen not just because of the short distance, 
	but also because the results highlight another conceptual point we stress 
	in this paper: 
	\begin{enumerate}
	\item[5)] Approaching total power capacity requires an increase in the complexity of 
	both the code design and the corresponding decoding algorithm 
	as communication distance is increased, or bit-error probability is lowered. 
	\end{enumerate}
	
	The results presented in this paper have some limitations. First, we only consider a limited set of coding strategies, 
	and while the results and models presented here extend easily to irregular LDPC constructions, they are not necessarily 
	applicable to all decoders. Second, modern transceivers~\cite{LeeBook} contain many other processing power sinks, 
	including analog-to-digital converters (ADCs), digital-to-analog converters (DACs), power amplifiers, modulation, and 
	equalizers, and the power requirements of each of these components can vary\footnote{For example, the \emph{resolution} of ADCs 
	used at the receiver may vary with the code choice by virtue of the fact that changing the \emph{rate} of the 
	code may require a change in signaling constellation (when channel bandwidth and 
	data-rate are fixed).} based on the coding strategy. While recent works have started to address fundamental 
	limits~\cite{Murmann1} and modeling~\cite{analogfrontendmodels} of power 
	consumption of system blocks from a mixed-signal circuit design perspective, tradeoffs 
	with \emph{code choice} of these components remain relatively unexplored. Hence, while analyzing 
	decoding power is a start, other system-level tradeoffs should be addressed in future work. It is also of great interest to 
	understand tradeoffs at a network level (see~\cite{JSAC11Paper}), where multiple transmitting-receiving pairs 
	are communicating in a shared wireless medium. In such situations, one cannot simply increase transmit 
	power to reduce decoding power: the resulting interference to other users needs to be accounted for as well. 
	
	The remainder of the paper is organized as follows. Section~\ref{sec:asymptotic} states 
	the assumptions and notation used in the paper. Sections~\ref{subsec:vlsilayoutmodel} 
	to \ref{subsec:moderncomp} introduce theoretical models of VLSI circuits and decoding energy.
	Preliminary results are stated in Section~\ref{sec:preliminarylemmas}, which are used to analyze decoding 
	energy in Sections~\ref{sec:nodemodelanalysis} and~\ref{sec:wireanalysis}, 
	in the context of the question illustrated in Fig.~\ref{fig:question}a) (obtaining the scaling behavior). Section~\ref{sec:finitelengthintro} 
	discusses circuit-simulation-based numerical models of decoding power, in the context of the 
	question illustrated in Fig.~\ref{fig:question}b). Section~\ref{sec:conclusion} concludes the paper.

\vspace{-0.2cm}
\section{System and VLSI models for asymptotic analysis}
\label{sec:asymptotic}
	Throughout this paper, we rely on Bachmann-Landau notation~\cite{knuth} 
	(i.e.~``big-O'' notation). We first state a preliminary definition that is needed in 
	order to state a precise definition of the big-O notation that we use in this paper. 
	\begin{definition}
	\label{def:rightsidedset}
	\normalfont $\mathcal{X} \subseteq \mathbb{R}$ is a right-sided set if $\forall x \in \mathcal{X}$, 
	$\exists y \in \mathcal{X}$ such that $y > x$. 
	\end{definition}
	
	Some examples of right-sided sets include $\mathbb{R}$, $\mathbb{N}$, 
	and intervals of the form $[a, \infty)$, where $a$ is a constant. We now state the 
	Bachmann-Landau notation for non-negative real-valued functions defined on right-sided sets\footnote{Bounded intervals that are open on the right such 
	as $(-1, 0)$ or $[0,5)$ are also right-sided sets. Definition~\ref{def:bigonotation} can still be applied to 
	functions restricted to such sets, but we will not consider such functions in this paper.}.	
	\begin{definition}
	\label{def:bigonotation}
	\normalfont Let $f: \mathcal{X} \rightarrow \mathbb{R}^{\geq 0}$ and $g: \mathcal{X} \rightarrow \mathbb{R}^{\geq 0}$ 
	be two non-negative real-valued functions, both defined on a right-sided set $\mathcal{X}$. We state
	\begin{enumerate}
	\setlength{\itemsep}{5pt}
	\item $f(x) = \mathcal{O}(g(x))$ if $\exists x_1\in\mathcal{X}$ and $c_{1}>0$ s.t.\\ $f(x)\leq c_{1}g(x)$, $\forall x \geq x_1$. 
	\item $f(x) = \Omega(g(x))$ if $\exists x_2\in\mathcal{X}$ and $c_{2}>0$ s.t.\\ $f(x)\geq c_{2}g(x)$, $\forall x \geq x_2$.
	\item $f(x) = \Theta(g(x))$ if $\exists x_3\in\mathcal{X}$ and $c_{4}\geq c_{3} > 0$ s.t.\\ $c_{3}g(x)\leq f(x)\leq c_4g(x)$, $\forall x \geq x_3$. 
	\end{enumerate} 
	\end{definition}	

	We will also need a Bachmann-Landau notation for \emph{two} variable 
	functions~\cite[Section 3.5]{brassardbratley}:
	\begin{definition}
	\label{def:bigonotationmult}
	\normalfont Let $u: \mathcal{X} \times \mathcal{Y} \rightarrow \mathbb{R}^{\geq 0}$ and 
	$v: \mathcal{X} \times \mathcal{Y} \rightarrow \mathbb{R}^{\geq 0}$ 
	be two non-negative real-valued functions, both defined on the Cartesian product of 
	two right-sided sets $\mathcal{X}$ and $\mathcal{Y}$. We state
	\begin{enumerate} 
	\setlength{\itemsep}{5pt}
	\item $u(x,y) = \mathcal{O}(v(x,y))$ if $\exists M\in\mathbb{R}$ and $c_{1}>0$ s.t.\\ $u(x,y) \leq c_1 v(x,y)$, $\forall x,y\geq M$.
	\item $u(x,y) = \Omega(v(x,y))$ if $\exists M\in \mathbb{R}$ and $c_{2}>0$ s.t.\\ $u(x,y)\geq c_2 v(x,y)$, $\forall x,y\geq M$.
	\item $u(x,y) = \Theta(v(x,y))$ if $\exists M \in \mathbb{R}$ and $c_{4}\geq c_{3}>0$ s.t.\\ $c_{3}v(x,y)\leq u(x,y)\leq c_{4}v(x,y)$, $\forall x,y\geq M$.
	\end{enumerate} 
	\end{definition}	

	We will often apply Definitions~\ref{def:bigonotation} and~\ref{def:bigonotationmult} in the limit as bit-error probability 
	$P_e \to 0$, where the definitions can be interpreted as applied to a function with an argument $\frac{1}{P_e}$ as it diverges 
	to $ \infty$. All logarithm functions $\log (\cdot)$ are natural logarithms unless otherwise stated.
	
	\vspace{-0.3cm}
	\subsection{Communication channel model}
	\label{subsec:channelmodel}
	We assume the communication between transmitter and receiver takes place over 
	an AWGN channel with fixed attenuation. The transmission strategy uses 
	BPSK modulation, and a $(d_v,d_c)$-regular binary LDPC code of design 
	rate $\rate = 1 - \frac{d_v}{d_c}$~\cite{urbankecapacity} 
	(which is assumed to equal the code rate). The blocklength of the code is denoted by $n$, and the number of source bits is denoted by $k=n \rate$. 
	The decoder performs a hard-decision on the observed channel outputs before starting the 
	decoding process, thereby first recovering noisy codeword bits 
	transmitted through a Binary Symmetric Channel (BSC) of 
	flip probability $p_0 = \mathbb{Q}\left(\sqrt{2 \frac{E_s}{N_0}}\right)$. Here, $E_s$ is the
	input energy per channel symbol and $\frac{N_0}{2}$ is the noise power.
	$\mathbb{Q} \left( \cdot \right)$ is the tail probability of the standard normal 
	distribution, $\mathbb{Q}(x) = \frac{1}{\sqrt{2 \pi}}\int_{x}^{\infty}e^{\frac{-u^2}{2}}\,du$.
	The transmit power $P_T$ is assumed to be proportional to $\frac{E_s}{N_0}$, modeling 
	fixed distance and constant attenuation wireless communication. Explicitly we assume 
	$\frac{E_s}{N_0} = \eta P_T$ for some constant $\eta > 0$. Using known bounds on the $\mathbb{Q}$-function~\cite{millsratio}, 
	$\frac{e^{-x^2/2}}{\sqrt{2\pi} (x+1/x)}\leq \mathbb{Q}(x)\leq \frac{e^{-x^2/2}}{\sqrt{2\pi} x}$ :
	\begin{equation}
	\label{eq:millsratioofficial} \frac{e^{-\eta P_T}}{\sqrt{4 \pi \eta P_T} + \sqrt{\frac{\pi}{\eta P_T}}} \leq p_0 \leq \frac{e^{-\eta P_T}}{\sqrt{4 \pi \eta P_T}}. 
	\end{equation}
	
	The focus of this paper is on the analysis of the ``total'' power required to communicate 
	on the above channel, as the target average bit-error probability $P_e \to 0$.  Our simplified notion of 
	total power is defined below.
	\begin{definition}
	\label{def:totalpower}
	\normalfont The total power, $P_\mathrm{total}$, consumed in communication across the channel 
	described in~\ref{subsec:channelmodel} is defined as
	\begin{equation}
		\label{eq:totalpowerdefinition} P_\mathrm{total} = P_T  + P_\mathrm{Dec},
	\end{equation}
	where $P_T$ and $P_\mathrm{Dec}$ are the power spent in transmission and decoding, respectively. 
	\end{definition}
	The channel model helps analyze the transmit power component in~\eqref{eq:totalpowerdefinition}, 
	but a model for decoding power is also needed. In the next section, we provide models and assumptions 
	for decoding algorithms and implementations that are used in the paper. We allow $P_T$ and $P_\mathrm{Dec}$ to be 
	chosen depending on $P_e$, $\eta$, and the coding strategy. Throughout the paper,
	the minimum total power for a strategy is denoted by $P_{\mathrm{total,min}}$ and the optimizing transmit 
	power by $P_T^*$.
	
	\vspace{-0.3cm}
	\subsection{Decoding algorithm assumptions}
	\label{subsec:decodingalg}
	The general theoretical results of this paper (Lemma~\ref{lemma:generalblocklength}, Theorem~\ref{thm:anyLDPClower}) 
	hold for any iterative message-passing decoding algorithm (and \emph{any} number of decoding iterations) 
	that satisfies ``symmetry conditions'' in~\cite[Def.~1]{urbankecapacity} (which allow us to assume that 
	all-zero codeword is transmitted). Thus, each node only operates on the messages it receives at its inputs. 
	We note that the sum-product algorithm~\cite{sumproduct}, the min-sum algorithm~\cite{minsum}, 
	Gallager's algorithms~\cite{gallagerthesis}, and most other message-passing decoders satisfy these assumptions. 
	For the constructive results of this paper (Corollary~\ref{cor:galanode}, Corollary~\ref{cor:galbnode},
	Theorem~\ref{thm:gallageratotalpower}, Theorem~\ref{thm:gallagebatotalpowerlower}) 
	we focus on the two decoding algorithms originally proposed in Gallager's 
	thesis~\cite{gallagerthesis}, that are now called ``Gallager-A'' and ``Gallager-B''~\cite{urbankecapacity}. 
	For these results, we will use density-evolution analysis~\cite{ModernCodingTheory} to analyze the 
	performance\footnote{In practice, decoding is often run for a larger number of iterations because at large 
	blocklengths, bit-error probability may still decay as the number of iterations increase. In that case, 
	density-evolution does not yield the correct bit-error probability, as it will vary based on the code 
	construction~\cite{errorflooremu}.}, for which we define the term ``independent iterations'' as follows:
	\begin{definition}
	\label{def:independentiterations}
	\normalfont An \textit{independent decoding iteration} is a decoding iteration in which
	messages received at a single variable or a check node are mutually independent.
	\end{definition}
	
	We will denote number of independent iterations\footnote{In our constructive results, we constrain the decoder to only 
	perform independent iterations. Thus, the number of independent iterations is the same as the number of iterations for those results, 
	but it emphasizes on the requirement on the code to ensure that the girth is sufficiently large.} that an algorithm runs as 
	$N_{\mathrm{iter}}$. This quantity is constrained by the \textit{girth}~\cite{urbankecapacity} of the code, defined as the 
	length of the shortest cycle in the Tanner graph of the code~\cite{Tanner} as follows: for a code with girth $g$, the maximum value of 
	$N_{\mathrm{iter}}$ is $\lfloor \frac{g-2}{4} \rfloor$.

	\vspace{-0.4cm}
	\subsection{VLSI model of decoding implementation}
	\label{subsec:vlsilayoutmodel}
	Theoretical models for analyzing area and energy costs of VLSI circuits were introduced several decades ago 
	in computer science. These include frameworks such as the Thompson~\cite{thompsonthesis} and 
	Brent-Kung~\cite{brentkung} models for circuit area and energy complexity (called the ``VLSI models''), and 
	Rent's rule~\cite{rentrule1,rentrule2}. Our model for the LDPC decoder implementation in this paper is an 
	adaptation of Thompson's model~\cite{thompsonthesis}, and it entails the following assumptions:
	\begin{enumerate}
	\item The VLSI circuit includes processing nodes which perform computations and  
	store data, and wires which connect them. The circuit is 
	placed on a square grid of horizontal and vertical wiring tracks of finite 
	width $\lambda > 0$, and contact squares of area $\lambda^2$ 
	at the overlaps of perpendicular tracks. 
	\item Neighboring parallel tracks are spaced apart by width $\lambda$. 
	\item Wires carry information bi-directionally. Distinct wires can only cross orthogonally at the contact squares.
	\item The layout is drawn in the plane. In other words, the model does not 
	allow for more than two metal layers for routing wires in the manner that modern 
	IC manufacturing processes do (see Section~\ref{subsubsec:multirouting}).
	\item The processing nodes in the circuit have finite memory and 
	are situated at the contact squares of the grid. They connect to wires routed along the grid. 
	\item Since wires are routed only horizontally and 
	vertically, any single contact has access to a maximum 
	of $4$ distinct wires. To accommodate higher-degree nodes, a 
	processing element requiring $x$ external connections 
	(for $x > 4$) can occupy a square of side-length $x \lambda$ on the 
	grid, with wires connecting to any side. No wires pass over the large square.
	\end{enumerate}
	
\begin{figure}[htbp]
   	\centering 
   	\includegraphics[width=\columnwidth]{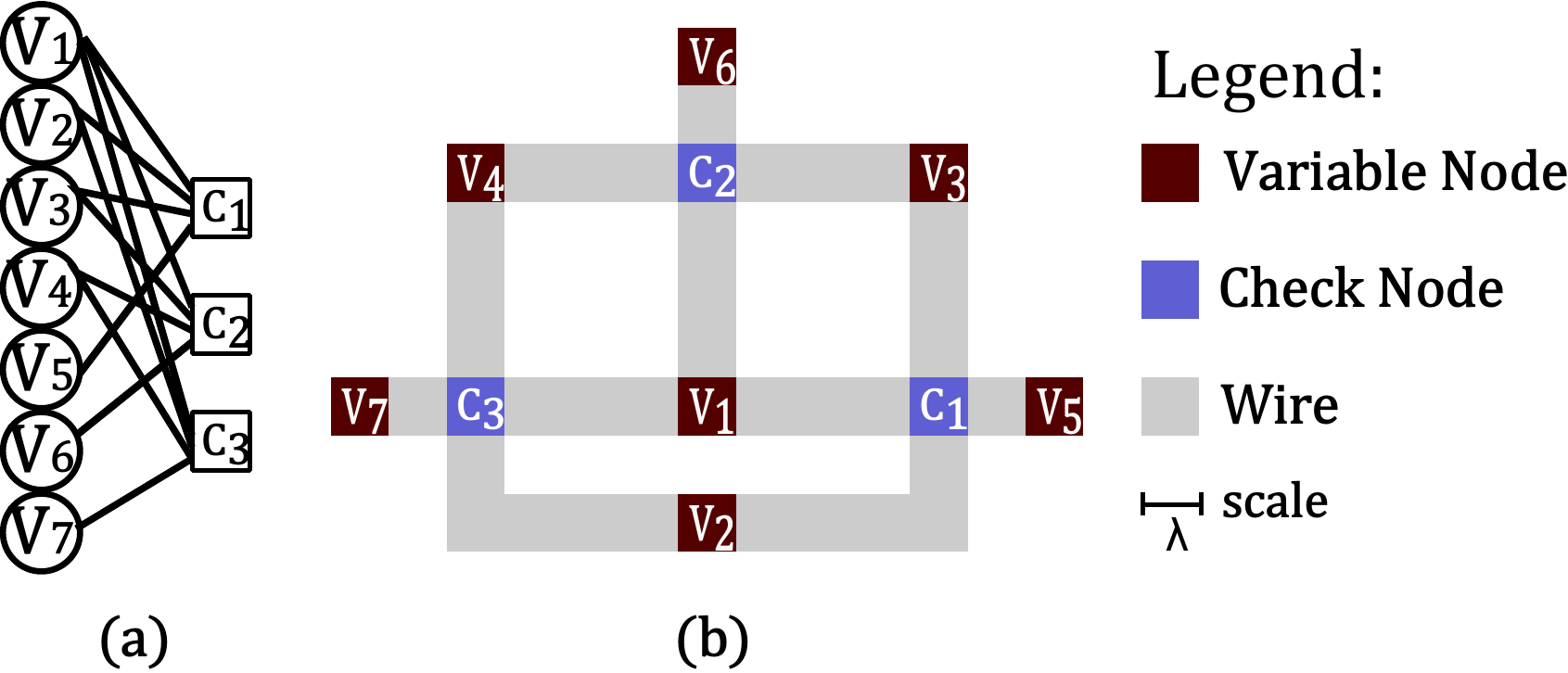} 
   	\caption{The Tanner graph (a) of a $(7,4)$-Hamming code and a fully parallel 
   	decoder (b) drawn according to Implementation Model ($\lambda$). Each vertex in 
	the Tanner graph corresponds to a processing node in the layout 
	and each edge in the Tanner graph corresponds to a wire connecting distinct nodes.}
   	\label{fig:Hamming}
  	\end{figure}

     	$\lambda$ models the minimum feature-size which is often used to describe IC 
	fabrication processes. We refer to this model as Implementation 
	Model ($\lambda$). The decoder is assumed to be implemented 
	in a ``fully-parallel'' manner~\cite{ldpcpowerreduction}, i.e. a processing node never 
	acts as more than one vertex in the Tanner graph~\cite{Tanner} of the code. 
	Each variable-node and check-node of an LDPC code is therefore represented by a
	distinct processing node in the decoding circuit. As an example, Fig.~\ref{fig:Hamming}(a) shows the 
	Tanner graph for a $(7,4)$-Hamming code and Fig.~\ref{fig:Hamming}(b) 
	shows a fully-parallel layout of a decoder for the same code.
	In Sections~\ref{subsec:nodemodel},~\ref{subsec:wiremodel} we will describe 
	two models\footnote{Both models can also be used simultaneously. However, for simplicity, we 
	present the results for the two models separately.} of energy consumption for the VLSI decoder. 
	
	\vspace{-0.35cm}
	\subsection{Time required for processing}
	\label{subsec:timecomputations}	
	In order to translate the model of~\ref{subsec:nodemodel} to a power model, we need the time required for computation 
	(the computation time is measured in seconds and is different from the number of
	algorithmic iterations). The computations are assumed to happen in clocked 
	iterations, with each iteration consisting of two steps: passing of messages from variable 
	to check nodes, and then from check to variable nodes. If the decoding 
	algorithm requires the exchange of multi-bit messages, we assume the message bits can be 
	passed using a single wire.
	
	We denote the decoding throughput (number of source bits 
	decoded per second) by $R_{\mathrm{data}}$. Because a batch of $k$ source 
	bits are processed in parallel, the time available for  
	processing is $T_{\mathrm{proc}}=\frac{k}{R_{\mathrm{data}}}$ seconds.
	
	\vspace{-0.2cm}
	\subsection{Processing node model of decoding power}
	\label{subsec:nodemodel}
	
	\begin{definition}[Node Model ($\xi_{\mathrm{node}}$)]
	\label{def:nodemodel}
	\normalfont The energy consumed in each variable or check node during one decoding
	iteration is $E_{\mathrm{node}}$. This constant can depend on $\lambda$, $d_v$ and $d_c$. 
  	The total number of nodes at the decoder\footnote{In practice, 
	many decoder implementations actually contain more than $n-k$ check nodes in order to break up small stopping-sets 
	in the code. However, we do not consider such decoders in this paper.} is $n_{\mathrm{nodes}} = n+(n-k) = 2n-k$. 
	The total energy consumed in $\tau_\mathrm{iter}$ decoding iterations is
	$E_{\mathrm{nodes}} = E_{\mathrm{node}} n_{\mathrm{nodes}} \tau_{\mathrm{iter}}$. 
	The decoding power is
	$P_{\mathrm{nodes}} = \frac{E_{\mathrm{nodes}}}{T_{\mathrm{proc}}} = \frac{E_{\mathrm{node}}(2n-k) \tau_{\mathrm{iter}}}{k} R_{\mathrm{data}} = \xi_{\mathrm{node}} 		
	\tau_{\mathrm{iter}}$.
	\end{definition}
	
	Here, $\xi_{\mathrm{node}} = E_{\mathrm{node}}\left(\frac{2}{R}-1\right)R_\mathrm{data} = \frac{E_\mathrm{node} \left( d_v + d_c \right)}{\left( d_c - d_v \right)}R_\mathrm{data}$. 	Note that $\tau_\mathrm{iter}$ need not be tied to $N_\mathrm{iter}$. 
	
	This model assumes that the entirety of the decoding energy is consumed in processing nodes, 
	and wires require no energy. In essence, this model is simply counting the number of operations 
	performed in the message-passing algorithm. The next energy model complements the node model 
	by accounting for energy consumed in wiring.
	
	\vspace{-0.2cm}
	\subsection{Message-passing wire model of decoding power}
	\label{subsec:wiremodel}
	\begin{definition}[Wire Model ($\xi_\mathrm{wire}$)]
	\label{def:wiremodel}
	\normalfont The decoding power is
	$P_\mathrm{wires}=  C_{\mathrm{unit-area}} A_\mathrm{wires} V_{supply}^{2} f_\mathrm{clock}$, where 
	$C_{\mathrm{unit-area}}$ is the capacitance per unit-area of a wire, $V_{supply}$ is the supply voltage of the circuit, 
	$f_\mathrm{clock}$ is the clock-frequency of the circuit, and $A_\mathrm{wires}$ is the total area 
	occupied by the wires in the circuit. The parameters $C_\mathrm{unit-area}$ and $V_{supply}$ are technology
	choices that may depend on $\lambda$, $d_v$, and $d_c$. The parameter 
	$f_\mathrm{clock}$ also may depend on $\lambda$, $d_v$, $d_c$, $R_{\mathrm{data}}$, and the decoding algorithm. 
	For simplicity, we write $\xi_\mathrm{wire} =  C_{\mathrm{unit-area}} V_{supply}^{2} f_\mathrm{clock}$ and
	$P_\mathrm{wires} =  \xi_{\mathrm{wire}} A_{\mathrm{wires}}$.
	\end{definition}	
	
	Wires in a circuit consume power whenever they are ``switched,'' i.e., 
	when the message along the wire changes its value\footnote{Switching consumes 
	energy because wires act as capacitors that need to be charged/discharged. If 
	voltage is maintained, little additional energy is spent.}. The probability of wire switching in a message-passing
	decoder depends on the statistics of the number of errors in the received word. These 
	statistics depend on the flip-probability of the channel, which is controlled by the transmit power. 
	Further, as decoding proceeds the messages also tend to stabilize, reducing switching and hence 
	the power consumed in the wires. Activity-factor~\cite{JanBook} could therefore be introduced in the Wire Model 
	via a multiplicative factor between $0$ and $1$ that depends on $P_T$, $\eta$, and the 
	decoding algorithm, but modeling it accurately would require a \emph{very} careful analysis.

	\vspace{-0.2cm}
	\subsection{On modern VLSI technologies and architectures}
	\label{subsec:moderncomp}
	The VLSI model of Section~\ref{subsec:vlsilayoutmodel} 
	and the assumptions made about the decoding architecture may seem
	pessimistic compared to the current state-of-the-art. However, in this 
	section we justify our choices by explaining how many of the architecture and 
	technology optimizations that are helpful in current practice have no impact on 
	the conclusions derived by our theoretical analysis. 
	
	\subsubsection{Multiple routing layers}
	\label{subsubsec:multirouting}
	Modern VLSI technologies allow for upwards of $10$ metal layers for routing wires~\cite{wikimetals}. 
	While this helps \emph{significantly} in reducing routing congestion in practice (i.e., at finite blocklengths 
	and non-vanishing error-probabilities), it has no impact on asymptotic bounds on total power. As proved 
	in~\cite[Pages 36-37]{thompsonthesis}, for a process with $L$ routing layers, the area occupied by wires 
	is at least $\frac{A_\mathrm{wires}}{L^2}$, where $A_\mathrm{wires}$ is the area occupied by the same circuit 
	when only one metal layer is used. As long as $L$ cannot grow with the number of vertices in the graph (it would 
	be very unrealistic to assume it can), it has only a constant impact on wiring area lower bounds 
	(see Lemmas~\ref{lem:triviallower},~\ref{lem:lowerboundwiringarea}) and \emph{no} impact (since one can always 
	restrict routing to a single layer) on upper bounds (see Lemma~\ref{lem:upperwirearea}). It will become apparent 
	later in the paper therefore, that multiple routing layers do not effect any of the theoretical results we derive.
	
	On the other hand, having multiple \emph{active} layers with fine-grained routing between layers \emph{can}  
	lead to asymptotic reductions in wiring area for some circuits~\cite{thompsonmultilayers}. However, as it relates to practice, this is 
	far beyond the reach of any commercial foundry in existence today. Methods for designing and fabricating 
	such circuits (which rely on emerging nanotechnologies and emerging non-volatile memories~\cite{n3xt}) are 
	only now starting to be considered in research settings.
	
	\subsubsection{Architectural optimizations}
	\label{subsubsec:serialized}
	Fully parallel, one clock-cycle per-iteration decoders are not commonly used in practice.
	Instead, serialization by dividing the number of physical nodes in the circuit by a constant factor and using 
	time-multiplexing to cut down on wiring is often performed~\cite{zhengyaJournal,ldpcpowerreduction}. 
	This also requires a corresponding multiplication for the clock-frequency $f_\mathrm{clock}$ of the circuit to 
	maintain the same data-rate. Recall, that dynamic power consumed in wires is proportional to 
	$C_\mathrm{wires} V_{supply}^{2} f_\mathrm{clock}$~\cite{JanBook}. While decrease in wire capacitance 
	may allow the supply to be scaled down (leading to a reduction in power) without compromising timing, 
	it is \emph{not possible} to scale it down indefinitely, since transistors have a nonzero subthreshold slope~\cite[Section 2]{subthresholdlimits}. 
	In other words, once a lower limit on supply voltage is reached, even if $C_\mathrm{wires}$ can be made to 
	decay on the order of $\frac{1}{n}$, one would no longer achieve power savings due to the corresponding increase in $f_\mathrm{clock}$. 
	Thus, behavior of total power in the large blocklength limit will remain unchanged. Such architectural optimizations 
	do however, have a big impact in practice (e.g., at finite-blocklengths) since changes in constants matter then.
	
	\subsubsection{Leakage power}
	\label{subsubsec:leakage}	
	Later in the paper (Sections~\ref{subsec:comparefundamental},~\ref{subsubsec:comparisonfundamentalwires}), 
	we will compare bounds on total power under the Node and Wire Models. It will turn out that the two models 
	lead to very different insights, and the Wire Model results appear far more pessimistic. Which model 
	then is closer to reality? It turns out that the Node Model is actually \emph{very optimistic}. It assumes that each
	node consumes only constant energy per-iteration, \emph{irrespective} of the clock period. From a circuit perspective, 
	this is equivalent to assuming that the power consumption inside nodes is entirely dynamic~\cite{JanBook}, as the energy per-iteration does not 
	increase with the clock-period. This is \emph{far} from the reality in modern VLSI technologies. 
	Transistors are not perfect switches~\cite{leakagepower}, and \emph{every} check-node and variable-node 
	will consume a constant amount of \emph{leakage power} while the decoder is on, 
	\emph{regardless} of clock period and switching activity. It is easy to see then, that even if the 
	transistor leakage is very small, the decoding power \emph{must} scale as $\Omega \left(n \right)$. For instance, 
	even if the architecture is highly serialized, there is still leakage in each of the $\Theta \left( n \right)$ sequential elements (e.g., flip-flops, latches,
	or RAM cells) needed to store messages. It will become apparent later in the paper 
	that this simple analysis is enough to establish identical conclusions to the lower bounds of 
	Theorems~\ref{thm:anyLDPClower},~\ref{thm:gallageratotalpower}, and~\ref{thm:gallagebatotalpowerlower}. 
	Thus, the asymptotics of total power under the Wire Model should be viewed as \emph{much} better predictions 
	of what would actually happen inside the circuit at infinite blocklengths. 

\section{Preliminary results}
\label{sec:preliminarylemmas}
\vspace{-0.1cm}
	In this section, we provide some preliminary results that will be useful in 
	Sections~\ref{sec:nodemodelanalysis} and~\ref{sec:wireanalysis}. These include general 
	bounds on the blocklength of regular-LDPC codes and bounds on the minimum number of independent 
	iterations needed for Gallager decoders to achieve a specific bit-error probability.
	
	\vspace{-0.2cm}
	\subsection{Blocklength analysis of regular-LDPC codes}
	\label{subsec:blockanalysis}
	\begin{lemma}
	\label{lem:n}
	For a given girth $g$ of a $(d_v,d_c)$-regular LDPC code, a lower bound on the blocklength $n$ is
	\begin{align}
	n &\geq \left[\left(d_v -1 \right) \left(d_c - 1 \right)\right]^{\lfloor \frac{g-2}{4} \rfloor},
	\end{align}
	and an upper bound on the blocklength is given by
	\begin{equation}
	n \leq 2(d_v+ d_c) d_vd_c(2d_vd_c+1)^{\frac{3}{4}g}.
	\end{equation}
	\end{lemma}
	
	\begin{IEEEproof}
	For the lower bound, see~\cite[Appendix I]{CISS11Paper}, and for the upper bound, see~\cite[Claim 2]{CISS11Paper}.
	\end{IEEEproof}	
	
	\begin{lemma}
	\label{lemma:generalblocklength}
	For a $(d_v, d_c)$-regular binary LDPC code decoded using any iterative message-passing decoding algorithm 
	for any number of iterations, the blocklength $n$ needed to achieve bit-error probability $P_{e}$ is 
	\begin{numcases}{n=}
	\Omega \left( \left(\frac{\left(\frac{d_v - 2}{d_v (d_v - 1)} \right)^{2} \log \frac{1}{P_e}}{(1 + 9 \pi) \eta P_{T}} \right)^{\frac{1 + \frac{\log(d_c - 1)}{\log(d_v - 1)}}{2}}\right)
	&$d_v \geq 3$. \nonumber\\
	\Omega \left( \left(\frac{1}{P_e} \right)^{\frac{1}{\eta P_{T} (1 + 9 \pi) \left(2 + \frac{2}{\log (d_c - 1)} \right)}} \right) &$d_v = 2$. \nonumber
	\end{numcases}
	Here, $\eta > 0 $ is the constant attenuation in the AWGN channel (see Section~\ref{subsec:channelmodel}).
	\end{lemma}
	
	\begin{IEEEproof}
	See Appendix~\ref{app:prooflemmageneral}.
	\end{IEEEproof}
	\noindent
	\begin{proofsketch} We use a technique for the finite-length analysis of LDPC codes 
	from~\cite{KoetterVontobel1}. First, the \emph{pairwise} error-probability for any iterative message-passing 
	decoder is lower bounded in terms of $n$, $d_v$, $d_c$, and $\eta P_T$ using an expression for the minimum
	\emph{pseudoweight} (see Appendix~\ref{app:prooflemmageneral} for definition) of the code. Next,
	due to a simple relationship between bit-error probability and pairwise error-probability for binary linear codes 
	over memoryless binary-input, output-symmetric channels, the bit-error probability can be lower bounded in terms of
	$n$, $d_v$, $d_c$, and $\eta P_T$. Finally, algebraic manipulations, an application of~\eqref{eq:millsratioofficial}, and 
	an application of Definition~\ref{def:bigonotationmult} complete the proof.
	\end{proofsketch}
	
	\vspace{-0.4cm}
	\subsection{Approximation analysis of Gallager decoding algorithms}
	\label{subsec:approximations}
	In this section, we bound the number of independent decoding iterations 
	required to attain a specific bit-error probability with Gallager decoders. These bounds are 
	used in Sections~\ref{sec:nodemodelanalysis},~\ref{subsec:totalpowerwires} to prove 
	achievability results for total power.
	
	\begin{lemma} 
	\label{lem:itera}
	The number of independent decoding iterations $N_{\mathrm{iter}}$ needed to attain 
	bit-error probability $P_e$ with Gallager-A decoding is
	\begin{numcases}{N_{\mathrm{iter}}=}
	\Theta \left( \log \frac{1}{P_e} \right)&\emph{if} $P_T$ is held constant. \nonumber \\
   	\Theta \left( \frac{\log \frac{1}{P_e}}{\eta P_T} \right)&\emph{if} $P_T$ is not held fixed. \nonumber 
	\end{numcases}
	Here, $\eta > 0$ is the constant attenuation in the AWGN channel (see Section~\ref{subsec:channelmodel}).  
	\end{lemma}
	
	\begin{IEEEproof}
	See Appendix~\ref{app:proofitera}.
	\end{IEEEproof}
	\noindent
	\begin{proofsketch} We first define (based on the decoding threshold over the BSC~\cite{urbankecapacity}) 
	appropriate right-sided sets for analyzing the asymptotics of $N_{\mathrm{iter}}$ as a function 
	of $\frac{1}{P_e}$ and $P_T$. Then, we apply a first-order Taylor expansion to the recurrence relation for 
	bit-error probability under independent iterations of Gallager-A decoding from~\cite[Eqn.~(6)]{urbankecapacity} 
	and carefully bound the approximation error. We then show that for small enough $P_e$ or large enough $P_T$, the
	approximation error can be bounded by a multiplicative factor between $\frac{1}{2}$ and $1$. After some 
	algebraic manipulations and an application of~\eqref{eq:millsratioofficial}, we apply 
	Definition~\ref{def:bigonotation} to establish the first case and Definition~\ref{def:bigonotationmult} to establish the second case.
	\end{proofsketch}
	
	\begin{lemma}
	\label{lem:iterb}
	The number of independent decoding iterations $N_{\mathrm{iter}}$ needed to attain bit-error probability 
	$P_e$ with a Gallager-B decoder with variable node degree $d_v \geq 4$ is given by
	\begin{numcases}{N_{\mathrm{iter}}=}
	\Theta \left( \log \log \frac{1}{P_e} \right)&\emph{if} $P_T$ is held constant. \nonumber \\
	\Theta \left( \frac{\log\frac{\log\frac{1}{P_e}}{\eta P_T}}{\log \frac{d_v - 1}{2}}\right)&\emph{if} $\lim_{P_e \to 0} \frac{P_T}{\log \frac{1}{P_e}} = 0$. \nonumber 
	\end{numcases}
	Here, $\eta > 0$ is the constant attenuation in the AWGN channel (see Section~\ref{subsec:channelmodel}).
	\end{lemma}
	
	\begin{IEEEproof}
	See Appendix~\ref{app:proofiterb}. Importantly, this holds only if $d_v \geq 4$, otherwise Gallager-A and Gallager-B are equivalent. 
	Note that in the second case, we assume $P_T$ is a function of $\frac{1}{P_e}$, so \emph{both} expressions should be interpreted 
	with Definition~\ref{def:bigonotation}. Further, little generality is lost by the necessary condition for the second case, since uncoded transmission
	requires transmit power $\Theta \left(\log \frac{1}{P_e} \right)$ (see~\eqref{eq:millsratioofficial}).
	\end{IEEEproof}
	\noindent
	\begin{proofsketch} We follow exactly the same steps as the proof of Lemma~\ref{lem:itera}, 
	but instead use a higher-order Taylor expansion of the recurrence relation for bit-error probability 
	under Gallager-B decoding from~\cite[Eqn. 4.15]{gallagerthesis}.
	\end{proofsketch}

\vspace{-0.25cm}
\section{Analysis of energy consumption in the node model}
\label{sec:nodemodelanalysis}
	In this section, we investigate the question: as $P_e \to 0$,
	how does the total power under the Node Model (see Section~\ref{subsec:nodemodel}) 
	scale when Gallager decoders (restricted to independent iterations) are used?
	
	\vspace{-0.4cm}
	\subsection{Total power analysis for Gallager-A decoding}
	\label{subsec:gallageranodepower}

	\begin{corollary}
	\label{cor:galanode}
	The optimal total power under Gallager-A decoding (restricted to independent iterations) in 
	the Node Model ($\xi_{\mathrm{node}}$) for a binary $(d_v, d_c)$-regular LDPC code is
	\begin{equation*}
		P_{\mathrm{total,min}}= \Theta \left(\sqrt{\log\frac{1}{P_e}}\right)
	\end{equation*}
	which is achieved by transmit power $P_T^*=\Theta \left(\sqrt{\log\frac{1}{P_e}}\right)$.
	\end{corollary}
	
	\begin{IEEEproof}
	Applying Lemma~\ref{lem:itera} to the Node Model, if $P_T$ is held constant even as $P_e \to 0$, the
	power consumed by decoding is $\Theta \left( \log \frac{1}{P_e} \right)$. Since $P_{T}$ is constant,
	the total power is also $P_{\mathrm{total,bdd\;P_T}} = \Theta \left( \log\frac{1}{P_e} \right)$. If instead 
	$P_T$ is allowed to grow arbitrarily, the total power is given by
	\begin{equation}
	P_{\mathrm{total}}= P_T + P_\mathrm{Dec} = \Theta \left(P_T +  \frac{\log\frac{1}{P_e}}{\eta P_T} \right).
	\end{equation}
	Thus, optimizing the scaling behavior of the total power over transmit power functions $P_T$
	\begin{equation}
	P_{\mathrm{total,min}}= \min_{P_T}  \Theta \left( P_T + \frac{\log\frac{1}{P_e}}{\eta P_T} \right) = \Theta \left(\sqrt{\log\frac{1}{P_e}} \right),
	\end{equation}
	with optimizing transmit power $P_T^* = \Theta \left(\sqrt{\log\frac{1}{P_e}}\right)$. 
	\end{IEEEproof}
	
	\subsection{Total power analysis for Gallager-B decoding}
	\label{subsec:gallagerbnodepower}

	\begin{corollary}
	\label{cor:galbnode}
	The optimal total power under Gallager-B decoding (restricted to independent iterations) in 
	the Node Model ($\xi_{\mathrm{node}}$) for a binary $(d_v, d_c)$-regular LDPC code is
	\begin{equation*}
		P_{\mathrm{total,min}}= \Theta \left(\log\log\frac{1}{P_e}\right),
	\end{equation*}
	which is achieved by transmit power $P_T^{*}=\Theta \left(1 \right)$.
	\end{corollary}
	
	\begin{IEEEproof}
	If $P_T$ satisfies the condition stated in the second case of Lemma~\ref{lem:iterb}, 
	the total power in the Node Model is
	\begin{equation}
	\label{eq:b} P_{\mathrm{total}} = P_T + P_\mathrm{Dec} = \Theta \left(P_T +\frac{\log\frac{\log\frac{1}{P_e}}{\eta P_T}}{\log \frac{d_v - 1}{2}} \right).
	\end{equation}
	Minimizing the scaling behavior of~\eqref{eq:b}, the optimizing transmit power is 
	$P_T^{*} = \Theta \left(1 \right)$. The optimal total power is then
	\begin{equation}
	P_{\mathrm{total,min}} = \Theta \left(\log\log\frac{1}{P_e}\right).
	\end{equation}
	In this case the optimizing transmit power is bounded even as $P_e \to 0$.
	\end{IEEEproof}
	
	\vspace{-0.45cm}
	\subsection{Comparison with fundamental limits}
	\label{subsec:comparefundamental}
	Can we reduce the asymptotic growth of total power under the Node Model via a better code or a more sophisticated 
	decoding algorithm? After all, we limited our attention to regular LDPCs and simple one-bit message-passing algorithms. 
	It was shown in~\cite{JSAC11Paper} that under the Node Model and a fully-parallelized decoding 
	implementation such as Implementation Model ($\lambda$), the optimal total power is 
	lower bounded by $\Omega \left( \log \log \frac{1}{P_e} \right)$, matching Corollary~\ref{cor:galbnode}. In 
	fact, using a code which performs close to Shannon capacity can even reduce efficiency for this
	strategy: if a capacity-approaching LDPC code is used instead of a regular LDPC code, the infinite-blocklength 
	performance under the Gallager-B decoding algorithm equals that of regular-LDPCs with Gallager-A decoding. 
	In other words, the bit-error probability decays only exponentially (and not doubly-exponentially) with the number 
	of iterations under Gallager-B decoding if degree-2 variable nodes are present~\cite{Lentmaier05}, 
	and~\cite{amincapacity} shows that degree-2 variable nodes are \emph{required} in order to achieve 
	capacity (the fraction of degree-2 variable nodes required to attain capacity is characterized in~\cite{amincapacity}). 
	Thus, rather than searching for an irregular code that approaches capacity, an engineer might be better off using 
	a simpler regular code that approaches fundamental limits on total power.

\vspace{-0.2cm}
\section{Analysis of energy consumption in the wire model}
\label{sec:wireanalysis}

\vspace{-0.2cm}
\subsection{Bounds on wiring area of decoders}
\label{subsec:wiringbounds}	
	To make use of the energy model of Section~\ref{subsec:wiremodel}, we 
	must characterize the total wiring area of the decoder. We rely on techniques for upper and lower 
	bounds on the total wire area obtained for different computations in~\cite{thompsonthesis,leiserson,lipton,leightonthesis}. 
	We first introduce some graph-theoretic concepts that will prove useful in obtaining similar bounds for our problem.
	
	\subsubsection{Lower bound on wiring area}
	\label{subsubsec:lowerboundwires}
	We first provide a trivial lower bound on the wiring area of the decoder for 
	any regular-LDPC code implemented in Implementation Model ($\lambda$).
	
	\begin{lemma}
	\label{lem:triviallower}
	For a $(d_v,d_c)$-regular LDPC code of blocklength $n$, the wiring area 
	$A_{\mathrm{wires}}$ under Implementation Model ($\lambda$) is
	\begin{equation*}
	A_{\mathrm{wires}} \geq \lambda^2 d_v n.
	\end{equation*}
	\end{lemma}
	\begin{IEEEproof}
	There are $d_v n$ wires. Each wire has width $\lambda$ and minimum length $\lambda$ (no two wires overlap completely).
	\end{IEEEproof} 
	
	In his thesis~\cite{leightonthesis}, Leighton utilizes the \emph{crossing number}
	(a property first defined by Tur\'{a}n~\cite{turan}) of a graph as a tool for 
	obtaining lower bounds on the wiring area of circuits. Crossing numbers continue to be of 
	interest to combinatorialists and graph-theorists, and many difficult problems on finding exact crossing 
	numbers or bounds for various families of graphs remain open~\cite{openproblems}. We use the following two 
	definitions to introduce this property.
	
	\begin{definition}[Graph Drawing]
	\label{def:drawing}
	\normalfont A drawing of a graph $\mathcal{G}$ is a representation of $\mathcal{G}$ in the plane
	such that each vertex of $\mathcal{G}$ is represented by a distinct point and each edge is represented
	by a distinct continuous arc connecting the corresponding points, which does not cross itself. 
	No edge passes through vertices other than its endpoints and no two edges are 
	overlapping for any nonzero length (they can only intersect at points).
	\end{definition}
	
	\theoremstyle{definition}
	\begin{definition}[Crossing Number]
	\label{def:crossingnumber}
 	\normalfont The crossing number of a graph $\mathcal{G}$, $cr(\mathcal{G})$, 
 	is the minimum number of edge-crossings over all possible drawings of $\mathcal{G}$. 
 	An edge-crossing is any point in the plane other than a vertex of $\mathcal{G}$ where a 
 	pair of edges intersects.
	\end{definition}
	
	For any graph $\mathcal{G}$ (e.g., the Tanner graph of an LDPC code), the wiring area of the 
	corresponding circuit under Implementation Model ($\lambda$) is lower bounded as
	$A_{\mathrm{wires}} \geq \lambda^2 cr(\mathcal{G})$. This is due to the fact that any VLSI 
	layout of the type described in Section~\ref{subsec:vlsilayoutmodel} can be mapped to a drawing 
	of $\mathcal{G}$ in the sense of Definition~\ref{def:drawing}, by simply replacing each processing node 
	with a point in the plane and replacing each wire by line segments connecting two points. Therefore, 
	the minimum number of wire crossings of any layout of $\mathcal{G}$ is $cr(\mathcal{G})$. Since 
	every crossing has area $\lambda^2$, the inequality follows. We now need lower bounds on the 
	crossing number of a computation graph. In this paper, we make use of the following result~\cite{pachspencertoth} 
	that improves on earlier results~\cite{probmethod,ajtai,leightonthesis} and allows us to tighten 
	Lemma~\ref{lem:triviallower} for some codes.
	\begin{theorem}[Pach, Spencer, T\'{o}th~\cite{pachspencertoth}]  
	\label{thm:pachspencertoth}
	Let $\mathcal{G} = \{V,E\}$ be a graph with girth $g > 2 \ell$ and $ \left| E \right| \geq 4 \left| V \right|$. 
	Then $cr \left( \mathcal{G} \right)$ 
	satisfies
	\begin{equation*}
	 cr \left( \mathcal{G} \right) \geq k_{\ell} \frac{\left| E \right|^{\ell+2}}{\left| V \right|^{\ell+1}},
	\end{equation*}
	where $k_{\ell} = \Omega \left( \frac{1}{\ell^2 2^{2 \ell + 3}} \right)$ \normalfont{\cite{szekely}}. 
	\end{theorem}
	
	We now obtain lower bounds on wiring area given a lower bound on the number 
	of independent iterations the code allows.	
	\begin{lemma}[Crossing Number Lower Bound on $A_{\mathrm{wires}}$]  
	\label{lem:lowerboundwiringarea}
	For a $(d_v,d_c)$-regular LDPC code that allows for 
	at least $\underline{N}_{\mathrm{iter}}$ independent decoding iterations, the 
	wiring area $A_{\mathrm{wires}}$ of a decoder in Implementation Model ($\lambda$) is
	\begin{numcases}{A_{\mathrm{wires}}=}
	\Omega \left( e^{\gamma \underline{N}_{\mathrm{iter}}} \right)& for any $d_v$, $d_c$ \nonumber \\
	\Omega \left( e^{\underline{N}_{\mathrm{iter}} \log \frac{2 d^2_v d^2_c}{(d_v + d_c)^{2}} } \right)& if $d_v d_c \geq 4 (d_v + d_c)$. \nonumber
	\end{numcases}
	Here, $\gamma \in \left[\log{\left[\left(d_v -1 \right)\left(d_c - 1\right)\right]},3 \log(2d_v d_c+1)\right]$ is a constant that depends on the code construction. 
	\end{lemma}	
	
	\begin{IEEEproof}
	Let $\mathcal{C}$ be a $(d_v,d_c)$-regular LDPC code that allows for \textit{at least} 
	$\underline{N}_{\mathrm{iter}}$ independent decoding iterations. Since the girth
	$g$ of $\mathcal{C}$ must then satisfy $\lfloor \frac{g-2}{4} \rfloor \geq \underline{N}_{\mathrm{iter}}$, $g > 4 \underline{N}_{\mathrm{iter}} - 2$.
	From Lemma~\ref{lem:n} then, the blocklength $n$ of the code $\mathcal{C}$ is
	\begin{equation*}
	 n = \Omega \left( e^{\gamma \underline{N}_{\mathrm{iter}}} \right),
	\end{equation*}
	where $\gamma \in \left[ \log \left( \left(d_v -1 \right) \left( d_c - 1 \right) \right),3\log(2d_v d_c+1) \right]$.
	And from Lemma~\ref{lem:triviallower} we then have $A_{\mathrm{wires}} = \Omega \left(e^{\gamma \underline{N}_{\mathrm{iter}}} \right)$.
	Now, assume $d_v d_c \geq 4 (d_v + d_c)$. This requires that $d_c > d_v \geq 5$. Let $V_{\mathcal{C}}$, $E_{\mathcal{C}}$ 
	denote the sets of vertices and edges in the Tanner graph of $\mathcal{C}$. The sizes are $ \left| E_{\mathcal{C}} \right| = n d_v$ and 
	$\left| V_{\mathcal{C}} \right| = n \left(1 + \frac{d_v}{d_c} \right)$. We then carry out the following algebra
	\begin{equation}
	\nonumber d_v d_c \geq 4 (d_v + d_c) \Rightarrow n d_v \geq 4 n \left(1 + \frac{d_v}{d_c} \right).
	\end{equation}
	Hence, $\left| E_{\mathcal{C}} \right| \geq 4 \left| V_{\mathcal{C}} \right|$. Using the fact that
	$g > 4 \underline{N}_{\mathrm{iter}} - 2$, we apply Theorem~\ref{thm:pachspencertoth}
	\begin{align}
	\nonumber A_{\mathrm{wires}} &= \Omega \left( \frac{\lambda^2}{\left( 2 \underline{N}_{\mathrm{iter}} - 1 \right)^2 4^{2 \underline{N}_{\mathrm{iter}} +\frac{1}{2}}} \frac{ \left( n d_v \right)^{2 \underline{N}_{\mathrm{iter}}+1}}{\left( n \left(1 + \frac{d_v}{d_c} \right) \right)^{2 \underline{N}_{\mathrm{iter}}}} \right)\\ 
	\label{eq:szekelybound} &= \Omega \left( \frac{\lambda^2 {\left( \frac{e^{\gamma}}{16} \right)}^{\underline{N}_{\mathrm{iter}}}}{\left( 2 \underline{N}_{\mathrm{iter}} - 1 \right)^{2}}\left( \frac{d_v d_c}{d_v + d_c}\right)^{2\underline{N}_{\mathrm{iter}}}\right).
	\end{align}
	Then, because $e^{\gamma} \geq (d_v - 1)(d_c - 1) = d_v d_c -(d_v + d_c) +1 \overset{d_v d_c \geq 4 (d_v + d_c)}{\geq} 3(d_v + d_c) +1$, and because $d_c > d_v \geq 5$, 
	we must have $e^{\gamma} \geq 34$. Substituting into~\eqref{eq:szekelybound},
	 \begin{align*}
	A_{\mathrm{wires}} &= \Omega \left( \frac{\lambda^2 {\left( \frac{34}{16}\right)}^{\underline{N}_{\mathrm{iter}}}}{\left( 2 \underline{N}_{\mathrm{iter}} - 1 \right)^{2}} \left( \frac{d_v d_c}{d_v + d_c}\right)^{2\underline{N}_{\mathrm{iter}}}\right)\\
	 &= \Omega \left( \lambda^2 2^{\underline{N}_{\mathrm{iter}}} \left( \frac{d_v d_c}{d_v + d_c}\right)^{2\underline{N}_{\mathrm{iter}}} \right),
	\end{align*}
	and changes-of-base complete the proof.
	\end{IEEEproof}

	\vspace{-0.1cm}
	\subsubsection{Upper bound on wiring area}
	\label{subsubsec:upperboundwires}
	Since the total circuit area is always an upper bound on 
	the area occupied by wires, we use an upper bound on the 
	circuit area to obtain the following upper bound on the wiring area 
	based on the \emph{maximum} number of independent iterations 
	that the code allows for.
	
	\begin{lemma}[Upper bound on $A_{\mathrm{wires}}$]  
	\label{lem:upperwirearea}
	For a $(d_v, d_c)$-regular LDPC code that allows for 
	no more than $\overline{N}_{\mathrm{iter}}$ independent decoding iterations, 
	the decoder wiring area $A_{\mathrm{wires}}$ is
	\begin{equation*}
	 A_{\mathrm{wires}} = \mathcal{O} \left( e^{2 \gamma \overline{N}_{\mathrm{iter}}} \right).
	\end{equation*}
	Here, $\gamma \in \left[\log{\left(\left(d_v -1 \right)\left(d_c - 1\right)\right)},3\log(2d_v d_c+1)\right]$ is a constant that depends on the code construction.
	\end{lemma}
	
\begin{figure}[htbp]
   	\centering
   	\includegraphics[width=\columnwidth]{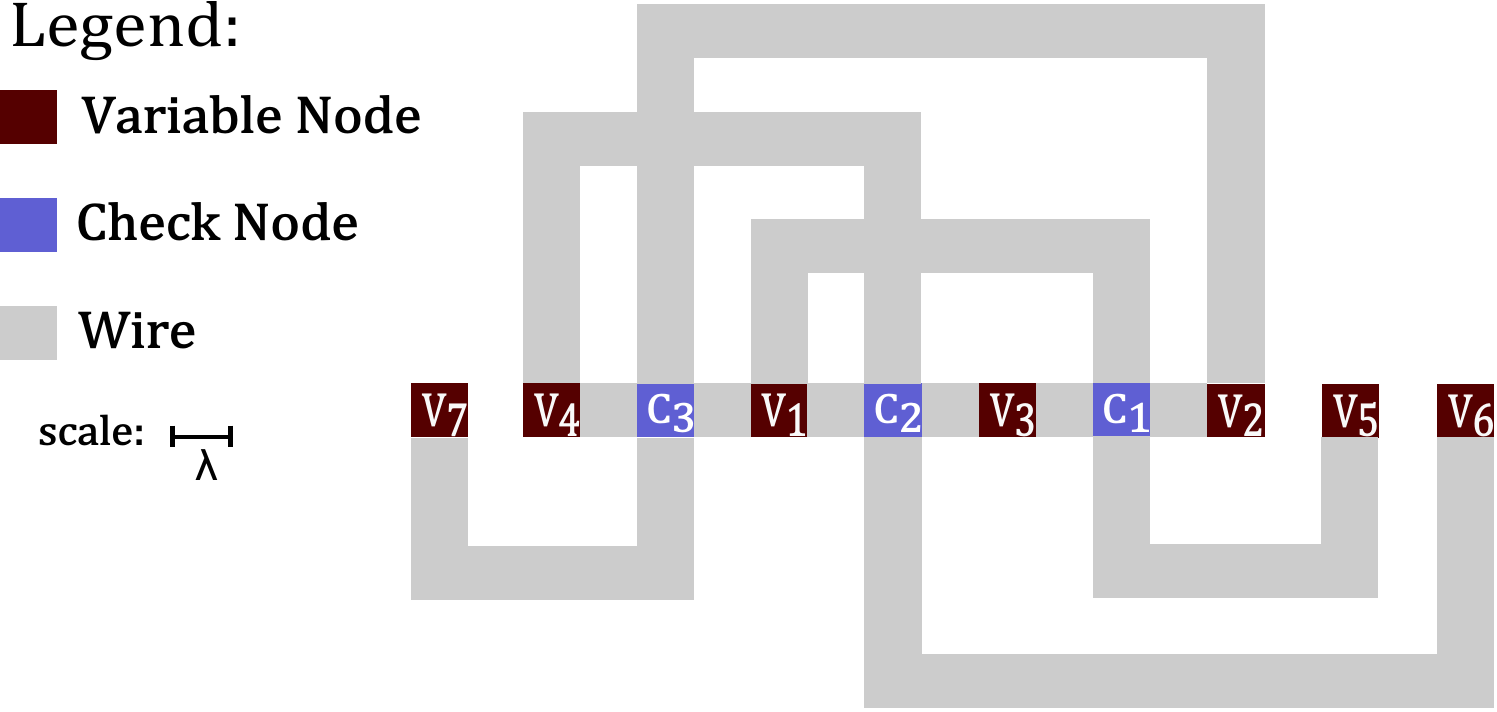} 
   	\caption{An example collinear layout for the same $(7,4)$ Hamming Code depicted in Fig.~\ref{fig:Hamming}.}
   	\label{fig:collinear}
	\end{figure}

	\begin{IEEEproof}
	Let $\mathcal{C}$ be a $(d_v,d_c)$-regular LDPC code that allows 
	for no more than $\overline{N}_{\mathrm{iter}}$ independent decoding 
	iterations. Since the girth $g$ of $\mathcal{C}$ must then satisfy 
	$\lfloor \frac{g-2}{4} \rfloor \leq \overline{N}_{\mathrm{iter}}$,
	\begin{equation*}
		g < 4 \overline{N}_{\mathrm{iter}} + 6.
	\end{equation*}
	From Lemma~\ref{lem:n}, the blocklength of any such 
	code can be upper bounded in the order of $\overline{N}_{\mathrm{iter}}$ as
	\begin{equation}
	 \label{eq:upperboundblock} n = \mathcal{O} \left( e^{\gamma \overline{N}_{\mathrm{iter}}} \right),
	\end{equation}
	where $\gamma \in \left[\log\left(\left(d_v -1 \right)\left(d_c - 1\right)\right),3\log(2d_v d_c+1)\right]$. Then, 
	consider a ``collinear'' VLSI layout~\cite{collinear} of the Tanner graph of 
	$\mathcal{C}$ which satisfies all the assumptions described in 
	Section~\ref{subsec:vlsilayoutmodel}. Arrange all variable-nodes 
	and check-nodes in the graph along a horizontal line, leaving 
	$\lambda$ spacing between consecutive nodes. The total length of 
	this arrangement is then $\mathcal{O}(n)$. Allocate a unique horizontal 
	wiring track for each of the $n d_v$ edges in the Tanner graph. Then, every 
	connection in the graph can be made with two vertical wires (one 
	from each endpoint) which connect to the opposite ends of the dedicated horizontal 
	track. The total height of this layout is then $\mathcal{O}(n)$, and the total area 
	is $\mathcal{O}(n^2)$. An example collinear layout is given in Fig.~\ref{fig:collinear}. 
	Substituting~\eqref{eq:upperboundblock} for $n$, we obtain the bound.
	\end{IEEEproof}
	
	We note that this upper bound is crude  
	since the $\mathcal{O} \left( \left| V \right|^2 \right)$ layout construction applies 
	for any graph $\mathcal{G} = \{V,E\}$ which satisfies  
	$\left| E \right| = \mathcal{O} \left( \left| V \right| \right)$. 
	A simple proof~\cite{leightonthesis} shows that one can create a layout of area
	$\mathcal{O} \left( \left( \left| V \right| + cr\left(\mathcal{G}\right) \right) \log \left( \left| V \right| + cr\left( \mathcal{G} \right) \right) \right)$
	for any graph. Thus, an algorithm for drawing semi-regular graphs which can be proven to yield 
	sub-quadratic (in $n$) crossing numbers would yield energy-efficient codes and decoders 
	with short wires.
	
	\vspace{-0.2cm}
	\subsection{Total power minimization for the wire model}
	\label{subsec:totalpowerwires}
	We now present analogues of results in Section~\ref{sec:nodemodelanalysis}, 
	where we instead consider decoding power described by the Wire Model of 
	Section~\ref{subsec:wiremodel}. We translate the wiring
	area bounds of Section~\ref{subsec:wiringbounds} to power bounds.
	
	\begin{theorem}[Asymptotic bounds on $P_{\mathrm{wires}}$]
	\label{thm:asymlbwire}
	Under Implementation Model ($\lambda$) and Wire Model ($\xi_{\mathrm{wire}}$), 
	the decoding power $P_\mathrm{wires}$ for a $(d_v, d_c)$-regular binary LDPC code that
	allows for \emph{exactly} $N_{\mathrm{iter}}$ independent iterations is bounded as
	\begin{numcases}{P_{\mathrm{wires}}=}
		\Omega \left( e^{\gamma N_{\mathrm{iter}}} \right)& for any $d_v$, $d_c$ \nonumber \\
		\Omega \left( e^{N_{\mathrm{iter}} \log \frac{2 d^2_v d^2_c}{(d_v + d_c)^{2}}} \right)& if $d_v d_c \geq 4 (d_v + d_c)$ \nonumber \\
		\mathcal{O} \left(e^{2\gamma N_{\mathrm{iter}}}\right)& for any $d_v$, $d_c$ \nonumber
	\end{numcases}
	where $\gamma \in \left[\log{\left(\left(d_v -1 \right)\left(d_c - 1\right)\right)},3\log(2d_v d_c+1)\right]$ is a constant that depends on the code construction..
	\end{theorem}
	\begin{IEEEproof}
	The result is a straightforward conclusion from Lemma~\ref{lem:lowerboundwiringarea} and Lemma~\ref{lem:upperwirearea} applied in Definition~\ref{def:wiremodel}.
	\end{IEEEproof}
	
	Next, we present a general lower bound on the scaling behavior of total power under 
	the Wire Model for any binary regular-LDPC code, decoded using any iterative message-passing 
	decoding algorithm, for any number of iterations.
	
	\begin{theorem}[Lower bound for regular-LDPCs]
	\label{thm:anyLDPClower}
	The optimal total power in the Wire Model ($\xi_{\mathrm{wire}}$) for a binary $(d_v, d_c)$-regular LDPC code 
	with any iterative message-passing decoding algorithm to achieve bit-error probability $P_{e}$ is
	\begin{equation*}
		P_{\mathrm{total,min}} = \Omega \left( \log^{\frac{1}{1 + \frac{2}{1 + \frac{\log (d_c - 1)}{\log (d_v - 1)} }}} \frac{1}{P_e} \right).
	\end{equation*}
	Further, if $P_{T}$ is held fixed as $P_{e} \to 0$ the total power diverges as $\Omega \left( \log^{y} \frac{1}{P_e} \right)$ 
	where $y > 1$, which dominates the power required by uncoded transmission.
	\end{theorem}
	\begin{IEEEproof}
	See Appendix~\ref{app:proofthmany}.
	\end{IEEEproof}	
	\noindent
	\begin{proofsketch} We first substitute the result of Lemma~\ref{lemma:generalblocklength} into Lemma~\ref{lem:triviallower}, 
	and then use the resulting lower bound on decoding power under the Wire Model in~\eqref{eq:totalpowerdefinition}. 
	Using simple calculus, we then derive the asymptotics of the transmit power function that minimizes the 
	total power, and plug it back into Lemma~\ref{lemma:generalblocklength} and~\eqref{eq:totalpowerdefinition} to obtain 
	the result.
	\end{proofsketch}
	
	\subsubsection{Gallager-A decoding}
	\label{subsubsec:gallagerawirepower}
	
	\begin{theorem}
	\label{thm:gallageratotalpower}
	The optimal total power under Gallager-A decoding (restricted to independent iterations) in the Wire 
	Model ($\xi_{\mathrm{wire}}$) for a binary $(d_v, d_c)$-regular LDPC code to achieve bit-error probability $P_e$ is 
	\begin{equation*}
		P_{\mathrm{total,min}} = \Theta \left(\frac{\frac{\gamma}{\eta}\log\frac{1}{P_e}}{\log \log \frac{1}{P_e}}\right),
	\end{equation*}
	Where $\eta > 0$ is the constant attenuation in the AWGN channel (Section~\ref{subsec:channelmodel}) 
	and $\gamma \in \left[\log{\left(\left(d_v -1 \right)\left(d_c - 1\right)\right)},3\log(2d_v d_c+1)\right]$ is 
	a constant that depends on the code construction. Further, if $P_T$ is held fixed as $P_e \to 0$, then
	total power diverges as $\Omega \left( \text{Poly} \left( \frac{1}{P_{e}} \right) \right)$, which is an
	exponential function of the power required by uncoded transmission.
	\end{theorem}
	
	\begin{IEEEproof}
		See Appendix~\ref{app:proofthma}.
	\end{IEEEproof}
	\noindent
	\begin{proofsketch} We first substitute the results of Lemma~\ref{lem:itera} into Theorem~\ref{thm:asymlbwire}, 
	and then plug in the resulting bounds on decoding power in~\eqref{eq:totalpowerdefinition}. 
	Using some calculus, we then derive the best-case and worst-case asymptotics of the transmit power 
	function that minimizes the total power, and show that there is at most a constant gap between the two. 
	We then plug the optimizing transmit power back into Lemma~\ref{lem:itera} and~\eqref{eq:totalpowerdefinition} to obtain the result.
	\end{proofsketch}
	
	\subsubsection{Gallager-B decoding}
	\label{subsubsec:gallagerbwirepower}

	\begin{theorem}
	\label{thm:gallagebatotalpowerlower}
	The optimal total power under Gallager-B decoding (restricted to independent iterations) in 
	the Wire Model ($\xi_{\mathrm{wire}}$) for a binary $(d_v, d_c)$-regular LDPC code to achieve 
	bit-error probability $P_e$ is bounded as 
	\begin{numcases}{P_{\mathrm{total,min}}=}
		\Omega \left(\log^\frac{2}{3} \frac{1}{P_e}\right)& $\frac{d_v d_c}{(d_v + d_c)} < 4$ \nonumber \\
		\Omega \left( \log^\frac{31}{40} \frac{1}{P_e} \right)& $\frac{d_v d_c}{(d_v + d_c)} \geq 4$ \nonumber \\
		\mathcal{O}\left(\log^{\frac{1}{1 + \frac{\log (d_v - 1) - \log 2}{6\log (2 d_v d_c + 1)}}} \frac{1}{P_e}\right)& any $d_v$, $d_c$ \nonumber
	\end{numcases}
	Further, if $P_T$ is held fixed as $P_e \to 0$, then total power diverges as $\Omega\left(\log^{2.48} \frac{1}{P_e}\right)$, 
	which is a super-quadratic function of the power required by uncoded transmission.
	\end{theorem}
	
	\begin{IEEEproof}
	See Appendix~\ref{app:proofthmb}.
	\end{IEEEproof}
	\noindent
	\begin{proofsketch} 
	We first substitute the results of Lemma~\ref{lem:iterb} into Theorem~\ref{thm:asymlbwire}, 
	and then plug in the resulting bounds on decoding power in~\eqref{eq:totalpowerdefinition}. We then use 
	algebraic manipulations to bound the exponents in Theorem~\ref{thm:asymlbwire}. Next, we use calculus 
	to derive the best-case and worst-case asymptotics of the transmit power function that minimizes the total power. 
	We then plug the optimizing transmit power into Lemma~\ref{lem:iterb} and~\eqref{eq:totalpowerdefinition} to obtain the results.
	\end{proofsketch}
	
	\vspace{-0.5cm}
	\subsection{Comparison with fundamental limits}
	\label{subsubsec:comparisonfundamentalwires}
	In~\cite{ISIT12Paper}, using a more pessimistic Wire Model\footnote{The Wire Model of~\cite{ISIT12Paper} 
	assumes the power is proportional to $A_{\mathrm{wires}}N_{\mathrm{iter}}$.
	Here it is assumed to be simply proportional to $A_{\mathrm{wires}}$.}, 
	it is shown that the total power required for any error-correcting code and any message-passing 
	decoding algorithm is fundamentally lower bounded by 
	$\Omega \left(\log^{\frac{1}{3}}\frac{1}{P^{blk}_e}\right)$, where $P^{blk}_e$ is the 
	$\emph{block}$-error probability. Theorem~\ref{thm:anyLDPClower} shows that
	regular-LDPC codes with iterative message-passing decoders cannot do better than
	$\Omega \left(\log^{\frac{1}{2}}\frac{1}{P_e}\right)$ where $P_e$ is bit-error probability, and 
	the exponent $\frac{1}{2}$ can only be obtained in the limit of large degrees and vanishing code-rate. Since 
	block-error probability exceeds bit-error probability, regular-LDPC codes do not achieve
	fundamental limits\footnote{Though, this may simply mean the fundamental limits~\cite{ISIT12Paper} are not tight.} 
	on total-power in the Wire Model. 
	
	Theorem~\ref{thm:gallageratotalpower} is the first constructive result that 
	shows that coding can (asymptotically) outperform uncoded transmission in total power 
	for the Wire Model. However, the gap in total power between the two is merely a multiplicative 
	factor of $\log \log \frac{1}{P_{e}}$. While Theorem~\ref{thm:gallagebatotalpowerlower} proves 
	that it is possible to increase the relative advantage of coding to a 
	fractional power of $\log \frac{1}{P_{e}}$, the difference between 
	the upper bound and the power for uncoded transmission is minuscule. The exponent of 
	$\log \frac{1}{P_e}$ in the upper bound is an increasing function of both $d_v$ and $d_c$, 
	approaching $1$ as either gets large. Since Gallager-B decoding requires $d_v \geq 4$, the smallest 
	exponent for regular LDPCs occurs when $d_v = 4$ and $d_c = 5$. The numerical value of the exponent 
	for these degrees is $\approx 0.98$, which suggests little order sense improvement over
	uncoded transmission. Hence, the wiring area at the decoder (particularly, 
	how much better it can be than the bound of Lemma~\ref{lem:upperwirearea})
  	is crucial in determining how much can be gained by using Gallager-B decoding instead 
	of uncoded transmission. Further discussion is provided in Section~\ref{sec:conclusion}.
	
	\vspace{-0.2cm}
	\section{Circuit simulation based numerical results}
	\label{sec:finitelengthintro}
	At reasonable bit-error probabilities (e.g., $10^{-5}$) and short distances 
	(e.g., less than five meters), asymptotic bounds cannot 
	provide precise answers on which codes to use. 
	For example, consider the following problem, shown graphically in 
	Fig.~\ref{fig:question}b).
	\begin{problem}
	\label{prob:1}
		\normalfont Suppose we want to design a point-to-point communication system that 
		operates over a given channel. We are given a target bit-error probability $P_e$, 
		communication distance $r$, and system data-rate 
		$R_{\mathrm{data}}$ that the link must operate at. Which code and corresponding 
		decoding algorithm minimize the total (i.e. transmit + decoding) power? 
	\end{problem}
		
	Since the bounds of Sections~\ref{sec:asymptotic}-\ref{sec:wireanalysis} 
	are derived as $P_{e} \to 0$, they may not be applicable to 
	many instances of Problem~\ref{prob:1}. In this section we
	therefore develop a methodology for rapidly exploring a space of
	codes and decoding algorithms to answer specific instances of Problem~\ref{prob:1}. 
	We focus on one-bit Gallager A and B~\cite{gallagerthesis} and two-bit~\cite{twobitdecoders} 
	decoding algorithms, restricting the number of algorithmic iterations 
	to $\lfloor \frac{g-2}{4} \rfloor$. Because of the effort required in 
	implementing or even simulating a single decoder in hardware, we 
	construct models\footnote{These models have been created in a open-access CMOS library~\cite{SynopsysLib} 
	and are online at~\cite{WebpageMod}.} for power consumed in decoding implementations of different algorithms 
	based on post-layout circuit simulations for simple check-node and variable-node circuits.
	The models developed attempt to capture detailed \emph{physical} aspects 
	(e.g., interconnect lengths and impedance parameters, 
	propagation delays, silicon area, and power-performance tradeoffs) 
	of implementations, in stark contrast with their theoretical counterparts
	of Sections~\ref{sec:asymptotic}-\ref{sec:wireanalysis}. In Section~\ref{subsec:optimizationexample},
	we use these models to investigate solutions to some instances of Problem~\ref{prob:1}. 

	\vspace{-0.2cm}
	\subsection{Note on channels and constellation size}
	\label{subsec:constellations}
	To answer Problem~\ref{prob:1}, additional physical assumptions about the channel
	(e.g., bandwidth, fading, path-loss, temperature, constellation size) are required in comparison 
	to the model of Section~\ref{subsec:channelmodel}. The channel is still assumed to be 
	AWGN with fixed attenuation. However, while Section~\ref{subsec:channelmodel} assumes BPSK 
	modulation for all transmissions, due to the introduction of a data-rate constraint 
	and fixed passband bandwidth $W$ (for fair comparison), the constellation size
	is required to vary based on the code rate. Explicitly,
	the transmission strategy is assumed to use either BPSK or square-QAM modulation, 
	mapping codeword bits to constellation symbols. We assume that if square-QAM modulation is used,
	the information bits are mapped onto the constellation signals using a two-dimensional Gray code
	as explained in~\cite[Section III]{QAM}. We assume the transmitter signals at a rate of $W$ symbols/s and 
	that the minimum square constellation size ($M$) satisfying the system data-rate requirement is chosen: $M$ is always  
	the smallest square of an even integer for which: 
		$$M \geq 2^{R_{\mathrm{data}}/ \left(W \times \rate \right)}.$$
	For calculating transmit power numbers, the thermal noise variance used is $\sigma_z^2=kTW$, 
	where $k$ is the Boltzmann constant ($1.38 \times 10^{-23}$ J/K), and $T$ is the temperature.
	The power is assumed to decay according to a power-law path-loss model $1/r^{\alpha}$, 
	where $\alpha$ is the path-loss coefficient. the received $\frac{E_b}{N_0}$ is obtained as a 
	function of the system and channel parameters:
	\begin{align}
		\frac{E_b}{N_0} &= \frac{P_T}{kTW \left(\frac{r}{\lambda}\right)^{\alpha}\log_2 (M)},
	\end{align}
	where $\lambda$ is the wavelength of transmission at center frequency 
	$f_c$ in Hz ($\lambda = 3 \times 10^8 / f_c$). The channel flip probability 
	for BPSK transmissions under this model is $p_0 = \mathbb{Q} \left( \sqrt{\frac{2 E_b}{N_0}} \right)$,
	and the channel flip probability for $M$-ary square QAM is~\cite[Section III.B]{QAM}:
	\begin{align}
		\nonumber p_0 &= \frac{1}{\log_2 (\sqrt{M})} \mathlarger{\sum}_{k=1}^{\log_2 (\sqrt{M})} \mathlarger{\sum}_{j=0}^{\left(1-\frac{1}{2^k}\right)
		\sqrt{M} -1} \bigg[ (-1)^{\big \lfloor \frac{j\times2^{k-1}}{\sqrt{M}} \big \rfloor}\\ 
		\nonumber &\times \left(2^{k-1} - \bigg \lfloor \frac{j\times2^{k-1}}{\sqrt{M}} + \frac{1}{2} \bigg \rfloor \right)\\ 
		\label{eq:QAM} &\times 2\mathbb{Q} \left ( (2j + 1)\sqrt{\frac{3 \frac{Eb}{N_0}\log_2 (M) }{ (M-1)}} \right) \bigg].
	\end{align}
	Also, note that the asymptotic bounds derived in  
	Sections~\ref{sec:asymptotic}-\ref{sec:wireanalysis} 
	\emph{remain unchanged}, even if we substitute $M$-ary QAM for BPSK as the 
	signaling constellation. This follows from the fact that the 
	RHS of equation~\eqref{eq:QAM} is a linear combination of
	$\mathbb{Q}\left( \cdot \right)$ functions with argument linearly 
	proportional to $\sqrt{\frac{E_b}{N_0}}$. Hence, even for $M$-ary QAM,
	$p_0 = \Theta \left( \frac{e^{-\phi P_T}}{\sqrt{4 \pi \phi P_T}} \right)$ for 
	some constant $\phi \neq \eta$ (see~\eqref{eq:millsratioofficial}). Since the difference is merely a constant, 
	the asymptotic analysis of Sections~\ref{sec:asymptotic}-\ref{sec:wireanalysis} holds.
	
	For the results presented in Section~\ref{subsec:optimizationexample}, 
	we assume the decoding throughput is required to be equal to 
	$R_{\mathrm{data}} = 7$ Gb/s. We assume a channel center frequency of 
	$f_c = 60$ GHz and bandwidth of $W = 7$ GHz. The temperature $T$ is 
	$300$ K. The distances considered are much larger than 
	the wavelength of transmission ($\approx$ $0.5$ cm) so the ``far-field approximation'' applies. 
	\vspace{-0.3cm}
	\subsection{Simulation-based models of LDPC decoders}
	\label{subsec:simmodels}
	Given a code, decoding algorithm, and desired data-rate, 
	calculating the required decoding power is a difficult task. 
	Even within the family of regular LDPC codes and specified decoding algorithms, 
	the decoder can be implemented in myriad ways. The choice of circuit 
	architecture, implementation technology, and even process-specific transistor 
	options can have a significant impact on the decoding power~\cite{ldpcpowerreduction,zhengyaJournal}. 
	A \emph{comprehensive solution} to 
	Problem~\ref{prob:1} requires optimization of total power over not just 
	\emph{super-exponentially} many codes and decoding algorithms, but also 
	\emph{all} decoder \emph{architectures}, \emph{implementation technologies}, 
	and \emph{process options}, which could be an impossibly hard problem. The models we present here are based on simulations of synchronous, 
	fully-parallel decoding architectures in a $32/28$nm CMOS process with a 
	high threshold voltage, and are used in Section~\ref{subsec:optimizationexample} to obtain insights on the nature 
	of optimal solutions. We believe that incorporating more models of this nature and performing the resulting optimization 
	could be a good approach to obtain low total power solutions. We now describe how the model is generated.

	\subsubsection{Initial post-layout simulations}
	\label{subsubsec:intialsims}
	Our models for arbitrary-blocklength LDPC decoders are constructed 
	based on circuit simulations using the Synopsys $32/28$nm high threshold voltage 
	CMOS process with $9$ metal-layers~\cite{SynopsysLib}. First, post-layout simulations 
	of check-node and variable-node circuits for one-bit and two-bit decoders 
	are performed. The physical area, power consumption, and critical-path 
	delays of the check-nodes and variable-nodes are used as the basis for
	our models. The CAD flow used is detailed in Appendix~\ref{app:cad}. 
	The next section details how these results are generalized to full decoders.

	\subsubsection{Physical model of LDPC decoding}
	\label{subsubsec:physicalmodel}
	Even within our imposed restrictions on the LDPC code degrees, 
	girth, and number of message-passing bits for decoding, constructing a 
	decoding power model that applies to all combinations of these code parameters 
	requires some assumptions:
	\begin{enumerate}
	\item[1.] Decoders operate at a fixed supply voltage (chosen as 
	$0.78$V: the minimum supply voltage of the timing libraries included with the standard-cell library).
	\item[2.] The code design space includes regular-LDPC codes with variable-node degrees
	$2 \leq d_v \leq 6$, check-node degrees $3 \leq d_c \leq 13$, and girths $6 \leq g \leq 10$.
	\item[3.] ``Minimum-Blocklength'' codes (found in~\cite{MERL}) are chosen for a 
	given $g,d_{v},d_{c}$. Hence the blocklength is expressed as a function of these parameters: 
	$n^{\mathrm{(min)}}_{g,d_{v},d_{c}}$.
	\item[4.] The decoding algorithm $a$, is chosen from the set $\{ A, B, T \}$, where $A$, $B$, $T$ 
	correspond to Gallager-A, Gallager-B, and Two-bit\footnote{With fixed decoding algorithm parameters 
	chosen as $C = 2$, $S = 2$, $W = 1$, for reasons explained in~\cite[Section II]{twobitdecoders}.}~\cite{twobitdecoders} 
	message-passing decoding algorithms, respectively. We use $\#_\mathrm{bits} \left( a \right)$ to refer to the number 
	of message bits used in algorithm $a$.
	\end{enumerate}
	
	We then model the minimum-achievable clock period $T_{\mathrm{CLK}}$, 
	and maximum-achievable decoding throughput $R_{\mathrm{Dec}}$ for each decoder 
	as functions of $a, g, d_{v}, d_{c}$:
	\begin{eqnarray}
	\nonumber T_{\mathrm{CLK}}\left(a, g, d_{v}, d_{c}\right) &=& T_{\mathrm{VN}}\left(a, d_{v}\right)\\
	\nonumber &+& 2T_{\mathrm{wire}}\left(a, g, d_v, d_c\right)\\
	\label{eq:delayeqn} &+& T_{\mathrm{CN}}\left(a, d_{c}\right)\\
	\label{eq:delayeqn2} R_{\mathrm{Dec}}\left(a, g, d_{v}, d_{c}\right) &=& \frac{n^{\mathrm{(min)}}_{g, d_{v}, d_{c}} \left(1 - \frac{d_{v}}{d_{c}} \right)}{\lfloor \frac{g-2}{4} \rfloor \times T_{\mathrm{CLK}}\left(a, g, d_{v}, d_{c}\right)}
	\end{eqnarray}
	 In~\eqref{eq:delayeqn}, $T_{\mathrm{VN}}\left(\cdot,\cdot \right)$ and 
	$T_{\mathrm{CN}}\left(\cdot,\cdot \right)$ are critical-path delays through
	variable and check nodes respectively and $T_{\mathrm{wire}}\left(\cdot,\cdot,\cdot,\cdot \right)$ 
	is the propagation delay through a single message-passing interconnect. 
	In essence,~\eqref{eq:delayeqn} formulates the critical-path delay for the decoder by summing up 
	the propagation delays of all logic stages traversed in a single decoding iteration. Details 
	for each component are given in Appendix~\ref{app:gatedelaymodels}. We model the decoding power as
	\begin{eqnarray}
	\nonumber P_{\mathrm{Dec}}\left(a, g, d_{v}, d_{c}\right) &=& n^{\mathrm{(min)}}_{g,d_{v},d_{c}} \bigg [P_{\mathrm{VN}}\left(a, d_{v}\right) + \frac{d_{v}P_{\mathrm{CN}}\left(a, d_{c}\right)}{d_{c}} \\ 
	\label{eq:pwreqn} &+& 2 d_v \#_\mathrm{bits} \left( a \right) \times P_{\mathrm{wire}}\left(a, g, d_{v}, d_{c}\right) \bigg].
	\end{eqnarray}
	In~\eqref{eq:pwreqn}, $P_{\mathrm{VN}}\left(\cdot,\cdot \right)$ and 
	$P_{\mathrm{CN}}\left(\cdot,\cdot \right)$ are the power consumed in individual 
	variable and check nodes respectively, and $P_{\mathrm{wire}}\left(\cdot,\cdot,\cdot,\cdot \right)$ 
	is the power consumed in a single message-passing interconnect. Note that~\eqref{eq:pwreqn} 
	is a sum of all power consumed in computations and wires of the decoder 
	(the coefficients in~\eqref{eq:pwreqn} count the number of occurrences of each power 
	sink in the decoder). The details of the node power models are given in Appendix~\ref{app:nodepowersim} 
	and the details of the wire power model are given in Appendix~\ref{app:wirepowersim}.
	\vspace{-0.1cm}
	\subsubsection{Satisfying the communication data-rate}
	\label{subsubsec:satisfydatarate}
	Fixing the supply voltage for a decoder and using the fastest possible 
	clock speed only allows for a single decoding throughput. Hence, parallelism in order 
	to meet the system data-rate requirement $R_{\mathrm{data}}$ in Problem~\ref{prob:1} is also 
	modeled. For example, two copies of a single decoder can be used in parallel. Together, 
	they provide twice the throughput, and require twice the power of a single decoder. 
	In the corresponding communication system architecture, two separate codewords are required to 
	be transmitted at twice the throughput of a single decoder, and a multiplexer 
	at the receiver must pass a separate codeword to each of the parallel decoders, which decode the two 
	codewords independently. Though making such a design choice in practice would introduce additional hardware 
	and a slight power consumption overhead, we ignore this cost in our analysis.
	\indent In cases where integer multiples of a single decoder's throughput do not exactly 
	reach $R_{\mathrm{data}}$, we first find the minimum number of parallel decoders, that when combined, 
	exceed the required throughput. Calling this minimum number of decoders $\mathcal{Q}$, we then 
	assume that the clock period of each of the parallel decoders is increased until the overall 
	throughput of the parallel combination is exactly $R_{\mathrm{data}}$. Explicitly, the formula to determine 
	this ``underclocked'' period $T_{u}$ is:
	\begin{align}
		\label{eq:explicittu} T_{u} &= \frac{\mathcal{Q} \times n^{\mathrm{(min)}}_{g, d_{v}, d_{c}} \left(1  - \frac{d_v}{d_c} \right)}{ \lfloor\frac{g-2}{4}\rfloor \times R_{\mathrm{data}}}.
	\end{align}
	Because the decoding power is modeled as inversely proportional 
	to the decoder clock period (see Appendices~\ref{app:nodepowersim}-\ref{app:wirepowersim}), 
	we multiply each individual decoder's power by the appropriate scaling factor 
	$\kappa = \frac{T_{\mathrm{CLK}} (a, g, d_v, d_c)}{T_{u}}$, and then multiply the result 
	by the number of parallel decoders to get the total power of the parallel combination:
	\begin{align}
		P_{\mathrm{parallel}} &= \mathcal{Q} \times P_{\mathrm{Dec}}(a, g, d_v, d_c) \times \kappa.
	\end{align}
	We substitute~\eqref{eq:delayeqn2},~\eqref{eq:explicittu}, and carry out some algebra to obtain:
	\begin{align}
		P_{\mathrm{parallel}} &= P_{\mathrm{Dec}} (a, g, d_v, d_c) \times \frac{R_{\mathrm{data}}}{R_{\mathrm{Dec}}(a, g, d_v, d_c)}.
	\end{align}
	Hence, we assume that any (throughput, power) pair that is a multiple of the 
	specifications of the original decoder can be achieved in this manner (with the obvious exception 
	of points that have negative throughput and power). Therefore, in our analysis in Section~\ref{subsec:optimizationexample}, 
	we assume the decoding throughput is exactly $R_{\mathrm{data}}$ and we use the decoding 
	power numbers obtained via this interpolation between the modeled points.

	\subsubsection{Comparing different coding strategies}
	\label{subsubsec:comparison}
	Now, given a subset of codes and decoders, how should a system designer 
	jointly choose a code and decoding algorithm to minimize the total system power?
	Within the channel model of Section~\ref{subsec:constellations}, consider 
	specific \emph{instances} of Problem~\ref{prob:1}: let path-loss coefficient $\alpha$ and 
	$R_{\mathrm{data}}$ be fixed. Then, for each choice of ($r$, $P_{e}$), we can compare the
	required total power for each combination of code and decoding algorithm
	modeled in Section~\ref{subsubsec:physicalmodel}, and find the minimizing combination.
	
	\vspace{-0.3cm}
	\subsection{Example: 60 GHz point-to-point communication}
	\label{subsec:optimizationexample}
\begin{figure}[htbp] %figure placement: here, top, bottom, or page
   	\centering
   	\includegraphics[width=0.9\columnwidth]{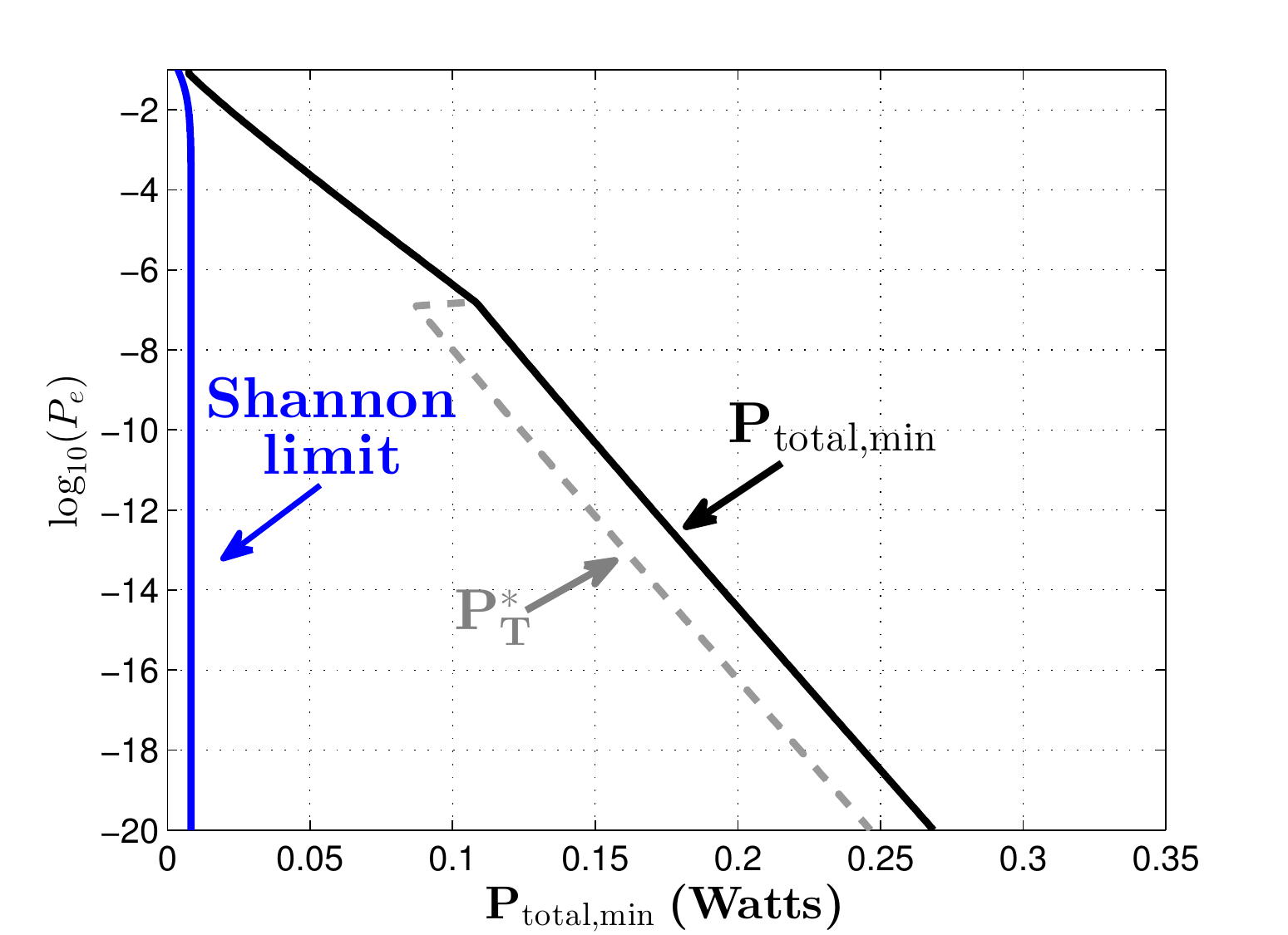} 
   	\caption{A plot of $\log_{10}(P_e)$ vs. minimum achievable total power for $\alpha = 3$ at a fixed distance of $r = 3.2$m. 
	The Shannon limit for the channel and the optimizing transmit power are also shown.}
   	\label{fig:totalpower28m}
	\end{figure}

	An example plot which shows the minimum achievable total power
	for different $P_e$ values at a fixed distance $r = 3.2$m and $\alpha = 3$ 
	is given in Fig.~\ref{fig:totalpower28m}. The plot also shows the 
	curve of the optimizing transmit power, $P^{*}_{T}$, and the Shannon-limit~\cite{ShannonOriginalPaper} 
	for the AWGN channel. The horizontal gap between the 
	optimizing $P_T$ curve and the total power curve 
	in Fig.~\ref{fig:totalpower28m} corresponds to the optimizing decoding power. 
	As $P_e$ decreases, this gap increases, indicating an 
	increase in the total power-minimizing decoder's complexity.
	
	\subsubsection{Joint optimization over code-decoder pairs}
	\label{subsubsec:jointoptim}
\begin{figure}[h]
	\begin{minipage}[c]{\columnwidth}
  	\vspace*{\fill}
  	\centering
  	\includegraphics[width=0.9\textwidth,left]{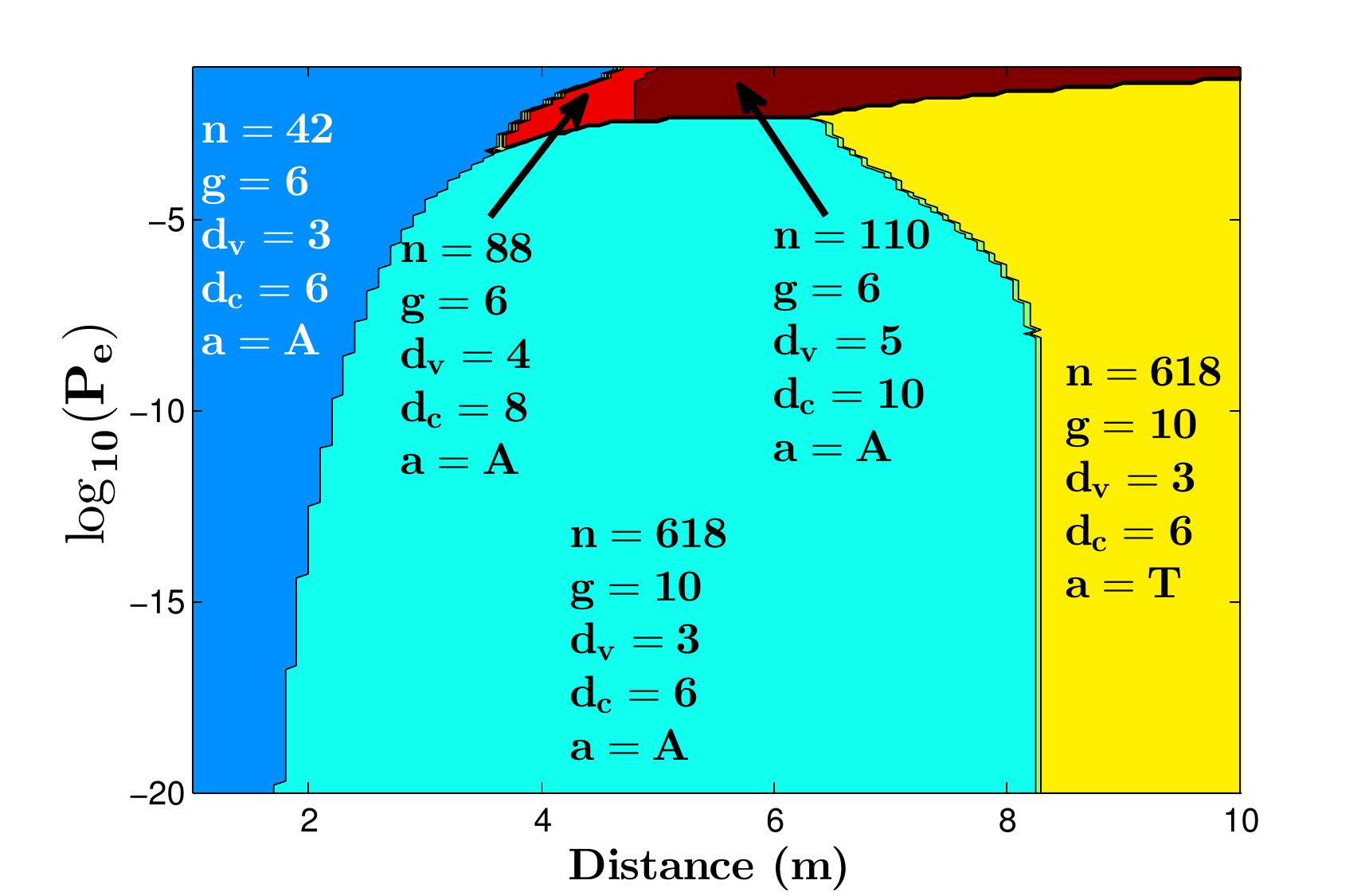}
  	\includegraphics[width=\textwidth]{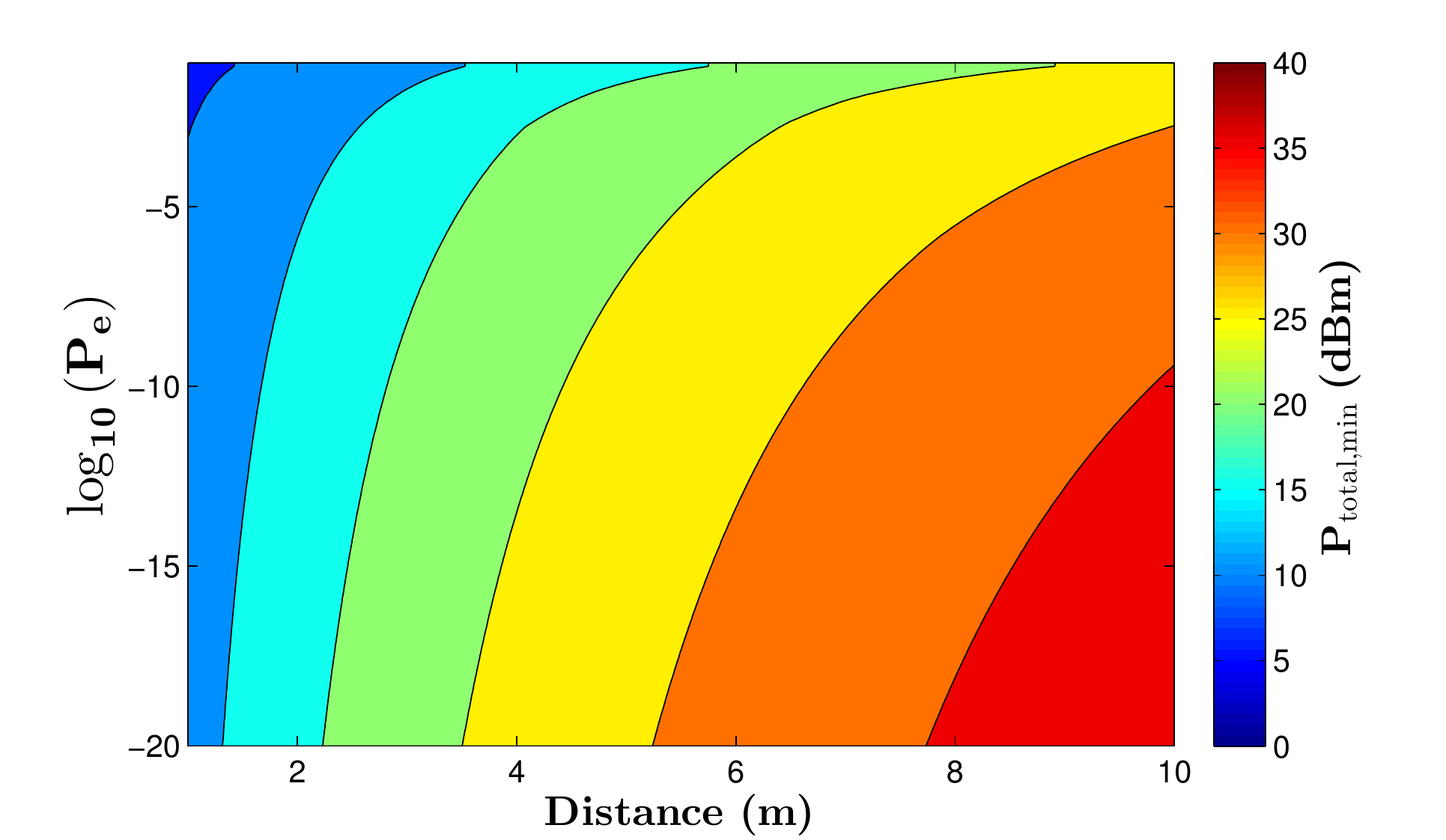}
	\end{minipage}
   	\caption{Contour plots of the optimizing code \& decoding algorithm 
   	 choice (top) and the minimum total power in dBm (bottom). For these plots, 
   	$\alpha = 3$. The contours in the top plots are labeled with blocklength $n$,
	 code girth $g$, VN degree $d_v$, CN degree $d_c$, and  
	 decoding algorithm $a$ of the optimizing code and decoder. To interpret the 
	 plots, one can choose any point in the ($r$, $P_e$) space and find the best
	 coding strategy (within the search space) in the top plot and
	 the required total power to implement it in the bottom plot. 
	 The plot is best viewed in color.}
   	\label{fig:nouncoded}
	\end{figure}

	The form of the total power curve varies with communication distance. For improved 
	understanding, we use two-dimensional contour plots in the ($r$, $P_e$) space to 
	evaluate choices of codes and decoders, as suggested by Fig.~\ref{fig:question}b). 
	An example is shown in Fig.~\ref{fig:nouncoded}, which compares code and decoding algorithm 
	choices for path-loss coefficient $\alpha=3$. In the top plot, the contours 
	represent regions in the ($r$, $P_e$) space where specific combinations minimize 
	total power, and in the bottom plot, regions in the ($r$, $P_e$) space are 
	divided based on the value of the minimum total power. The best 
	choices for these instances of Problem~\ref{prob:1} turn out to be rate $\frac{1}{2}$ codes. 
	Lower rate codes require large constellations for a $7$ Gb/s data-rate, thus requiring
	large transmit power for the same $p_{0}$, and higher rate codes require 
	larger decoding power due to increased complexity and size of higher degree
	nodes. Some tradeoffs between total power
	and code and decoder complexity can also be observed in 
	Fig.~\ref{fig:nouncoded}: to minimize total power, algorithm complexity 
	$a$ should increase with $r$ and code girth $g$ should increase 
	with decreasing $P_e$.

\begin{figure}[h]
   	\centering
	\begin{minipage}[c]{\columnwidth}
  	\vspace*{\fill}
  	\centering
  	\includegraphics[width=0.9\textwidth,left]{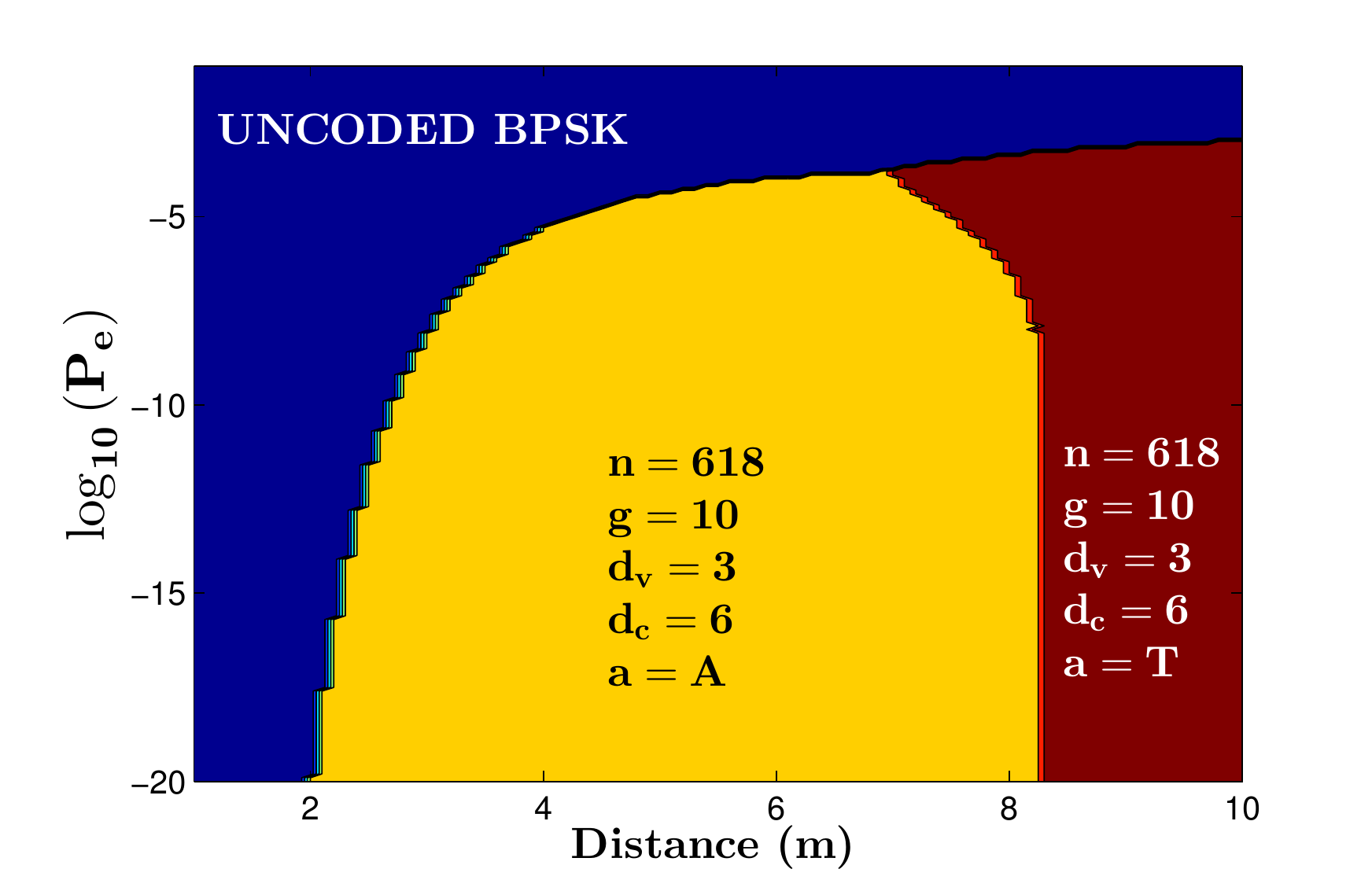}
  	\includegraphics[width=\textwidth]{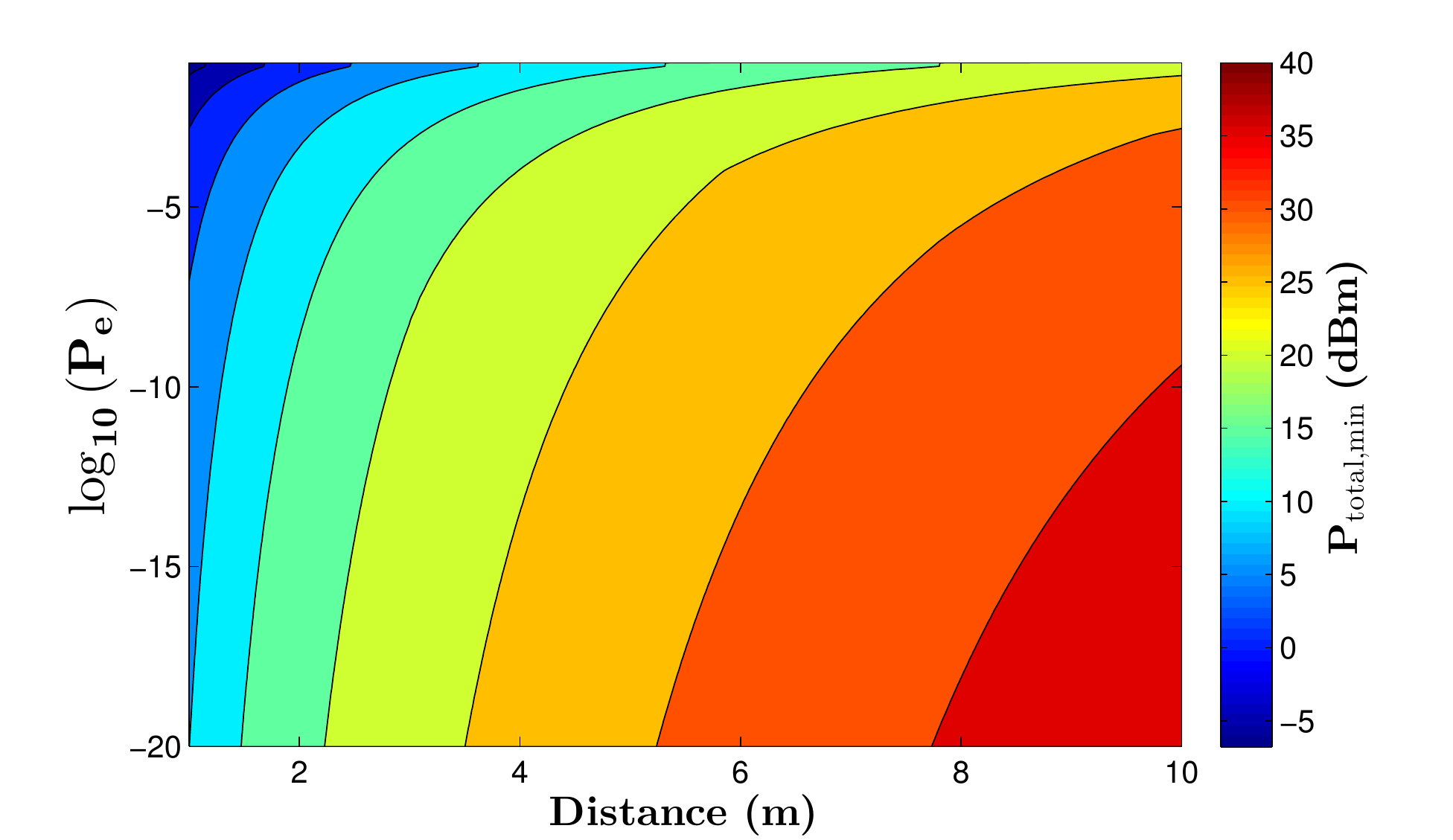}
	\end{minipage}
   	\caption{Contour plots of the optimizing code \& decoding algorithm 
   	 choice (top) and the minimum total power in dBm (bottom), including uncoded 
	 transmission in the optimization space. For these plots, $\alpha = 3$. The contours 
	 in the top plots are labeled with blocklength $n$, code girth $g$, VN degree 
	 $d_v$, CN degree $d_c$, and decoding algorithm $a$ of the optimizing code 
	 and decoder. To interpret the plots, one can choose any point in the ($r$, $P_e$) space 
	 and find the best coding strategy (within the search space) in the top plot and
	 the required total power to implement it in the bottom plot. 
	 The plot is best viewed in color.}
   	\label{fig:withuncoded}
	\end{figure}

	How does the inclusion of uncoded transmission as a possible 
	strategy change the picture? Contour plots with uncoded transmission included are 
	given in Fig.~\ref{fig:withuncoded}. Comparing Fig.~\ref{fig:withuncoded} with 
	Fig.~\ref{fig:nouncoded}, we see that when uncoded transmission is included, 
	it overtakes areas in the ($r$, $P_e$) space where $P_e$ is high 
	and $r$ is very small. However, Fig.~\ref{fig:withuncoded} suggests 
	that simple codes and decoders can still outperform uncoded transmission 
	at reasonably low $P_{e}$ and distances of several meters or more.
	
\vspace{-0.2cm}
\section{Conclusions and discussions}
\label{sec:conclusion}
	In this work, we performed asymptotic analysis of the total (transmit + decoding) power 
	for regular-LDPC codes with iterative-message passing decoders. While these codes (with Gallager-B decoding) 
	can achieve fundamental limits in the Node Model~\cite{JSAC11Paper}, they are unable to do so for the 
	Wire Model~\cite{ISIT12Paper}. This suggests that measuring complexity of decoding by simply counting the number of 
	operations (e.g.,~\cite{viterbi,JacobsBerlekamp,polarcodes}) is insufficient for understanding system-level power consumption. 
	In fact, for the Wire Model, even achieving order-sense advantage over \emph{uncoded transmission} requires
	that both transmit and decoding power diverge to $\infty$ as $P_{e} \to 0$, which calls into 
	question the assumption that one should fix the transmit power and operate near the Shannon capacity 
	in order communicate reliably at a low power cost. However, this analysis also established a result of intellectual interest: 
	that regular-LDPC codes \textit{can} achieve an order-sense improvement in total power over uncoded transmission as 
	bit-error probability tends to zero. This question only arises from the total power perspective adopted in this work, and 
	it suggests that these results are only scratching the surface of a deeper theory in this direction.
	
	To establish some constructive results, we analyzed two strategies where the number of decoding 
	iterations is dictated by the girth of the code. Although this is convenient for proving
	asymptotic upper bounds on total power, this is rarely followed in practice. Typically, combinatorial 
	properties of the code construction are analyzed and simulations are performed~\cite{errorflooremu} 
	in order to discern the error-probability behavior in decoding iterations beyond the girth-limit. However, we 
	are not sure if better asymptotics for total power can be achieved by merely adding these additional iterations.
	
	Our work highlights an important question that has received little attention in coding theory literature: 
	design of codes that have good performance while maintaining small wiring area
	(see~\cite{MansourShanbhag,Mohiyuddin,ThorpeISIT} for some heuristic approaches to generating 
	Tanner graphs with low wiring complexity). For wire power consumption, there is a significant gap between the bounds 
	on power consumed by regular-LDPC codes and iterative message-passing decoders 
	derived here, and the fundamental limits derived in~\cite{ISIT12Paper}. Nevertheless, even 
	though regular-LDPC codes might not achieve these fundamental limits 
	(and the fundamental limits themselves may not be tight), it is important to investigate wiring complexity
	of other coding families, such as Polar codes~\cite{polarcodes} and Turbo codes~\cite{turbocodes}.
	 
	Recent work of Blake and Kschischang~\cite{BlakeKschischang2} studied the limiting bisection-width~\cite{thompsonthesis} 
	of sequences of bipartite graphs with the size of the left-partite set tending to infinity, when the limiting degree-distributions 
	of the left and right partite sets satisfy a certain sufficient condition~\cite[Theorem 1]{BlakeKschischang2}. It is shown that
	when sequences satisfying this condition are generated by a standard uniform random configuration model 
	(see~\cite[Section IV]{BlakeKschischang2} for definition), the resulting graphs have a super-linear (in the number of vertices) 
	bisection-width in the limit of the sequence with probability $1$. In Corollary 2 and Section IV. A of~\cite{BlakeKschischang2}, 
	the authors show that the Tanner graphs of all capacity-approaching LDPC sequences as well as some regular-LDPC 
	sequences generated using this method will satisfy the sufficient conditions. A super-linear bisection width for a graph 
	implies that the area of wires in the corresponding VLSI circuit must scale at least quadratically in the number of 
	vertices~\cite{thompsonthesis}. If using the decoding strategy of Theorem~\ref{thm:gallagebatotalpowerlower} then, 
	such sequences of codes will have minimum total power that is $\Theta \left( \log^{m} \frac{1}{P_e} \right)$, where $0.97 < m < 1$, 
	providing little order-sense improvement over uncoded transmission. 
	
	The authors of~\cite{BlakeKschischang2} point out the fact that their result does not rule out the possibility that there 
	may exist a zero-measure (asymptotically in $n$) subset of codes that has sub-quadratic wiring area. One could try to extend 
	the bisection-width\footnote{The crossing number $cr \left( \mathcal{G} \right)$ and bisection width 
	$bw \left( \mathcal{G} \right)$ of a bounded-degree graph $\mathcal{G} = \{V, E\}$ are related by the inequality 
	$cr \left( \mathcal{G} \right) + \Theta \left( \left| V \right| \right) = \Omega \left( bw^{2} \left( \mathcal{G} \right) \right)$~\cite[Theorem 2.1]{crossingplanars}.} 
	approach of~\cite{BlakeKschischang2} to establish a negative result (i.e., prove that
	no such zero-measure set exists). To establish a positive result, one could try the open problem 
	mentioned at the end of Section~\ref{subsubsec:upperboundwires}, namely, construct a graph-drawing algorithm
	that yields sub-quadratic crossing numbers for (even some classes of) semi-regular graphs. In any case, a 
	proof is needed and heuristics such as those used in~\cite{ThorpeISIT} (even if they work well in practice) 
	cannot establish guarantees.

	The simulation-based estimates of decoding power presented in 
	Section~\ref{sec:finitelengthintro} confirm that coding can be useful for
	minimizing total power, even at short-distances. For instance, they predict that 
	regular-LDPC codes with simple message-passing decoders can achieve lower 
	bit-error probabilities than uncoded transmission in short distance settings, while still 
	consuming the same total power (even at distances as low as $2$ meters). However, 
	in these regimes, it is possible that ``classical'' algebraic codes (e.g., Hamming or 
	Reed-Solomon codes~\cite{LinCostello}) might be even more efficient, hence, they need to be 
	examined as well. 
	
	Finally, the results of Section~\ref{sec:finitelengthintro} point to a new problem, 
	that of ``energy-adaptive codes''. The suggestion from these results is that the code should be adapted 
	to changing error-probabilities and distances. Can a single code, with a single piece of reconfigurable 
	decoding hardware, enable adaptation of transmit and circuit power to minimize total energy? Indeed, some 
	follow-up work~\cite{HaewonPaper} indicates this is possible, and it could be a promising direction for future work.
	
\vspace{-0.2cm}
\section*{Acknowledgements}
\label{sec:ack}
	{\small This work was supported in part 
	by the NSF Center for Science of Information (CSoI) NSF CCF-0939370, 
	as well as a seed grant on ``Understanding Information-Energy Interactions'' 
	from the NSF CSoI. This work is also supported in part by Systems on Nanoscale 
	Information fabriCs (SONIC), one of the six SRC STARnet Centers, sponsored by 
	MARCO and DARPA. Grover's research supported by NSF CCF-1350314 
	(NSF CAREER), and NSF ECCS-1343324 (NSF EARS) awards. 
	We thank Anant Sahai, Subhasish Mitra, Yang Wen, Haewon Jeong, and Jianying Luo 
	for stimulating discussions, and Jose Moura for code constructions that we started our 
	simulations with. We thank the students, faculty, staff and sponsors of the Berkeley 
	Wireless Research Center and Wireless Foundations group at 
	Berkeley. In particular, Brian Richards assisted with the simulation 
	flow and Tsung-Te Liu advised us on modeling circuits. Finally, we thank the 
	anonymous reviewers whose comments helped us greatly in improving the manuscript.}

\appendices{}
\setcounter{section}{0}
\renewcommand\thesection{\Alph{section}}
	\vspace{-0.2cm}
	\section{Proof of Lemma~\ref{lemma:generalblocklength}}
	\label{app:prooflemmageneral}
	\allowdisplaybreaks
	\begin{proofof}{Lemma~\ref{lemma:generalblocklength}}
	First, note that the $\Omega \left( \cdot \right)$ expression in Lemma~\ref{lemma:generalblocklength} 
	contains two variables: $\frac{1}{P_e}$ and $P_{T}$. We analyze blocklength as a function 
	$n: [2,\infty) \times \mathbb{R}^{\geq 0} \to \mathbb{R}^{\geq 1}$ (Definition~\ref{def:bigonotationmult}). 
	First consider the case where $d_c > d_v \geq 3$. Because the codes considered are binary and linear, the 
	channel is memoryless, binary-input, and output-symmetric, and the decoding computations are symmetric 
	with respect to codewords, we can assume without loss of generality that the all-zero codeword was transmitted~\cite{urbankecapacity},~\cite[Page 22]{KoetterVontobel1}.
	In~\cite[Page 37]{KoetterVontobel1} it is shown that for \emph{any} binary regular-LDPC code with $d_c > d_v \geq 3$ 
	used to transmit over an AWGN channel, the probability that any iterative message-passing decoder incorrectly decides on 
	pseudo-codeword\footnote{Defined in~\cite{KoetterVontobel1} as an error-pattern for a given code $\mathcal{C}$, such that
	the lifting of the error-pattern is a codeword of the binary code corresponding to some finite graph-cover of the Tanner graph of $\mathcal{C}$. 
	It is explained in~\cite{KoetterVontobel1} that no ``locally operating'' iterative message-passing decoding algorithm 
	(``locally operating'' subsumes all algorithms satisfying the assumptions of Section~\ref{subsec:decodingalg}) can distinguish 
	between codewords and pseudo-codewords.} $\mathbf{\omega}$ when the all-zero codeword was transmitted is
	\begin{equation}
		\label{eq:pairwise} P_{0 \to \mathbf{\omega}} \geq \mathbb{Q} \left(\sqrt{2\frac{E_s}{N_0} w_{p}^{\mathrm{AWGN}} (\mathbf{\omega})} \right),
	\end{equation}
	where $w_{p}^{\mathrm{AWGN}} (\mathbf{\omega})$ is called the AWGN pseudoweight 
	of $\mathbf{\omega}$ and is defined~\cite[Definition 31]{KoetterVontobel1} as
	\begin{numcases}{w_{p}^{\mathrm{AWGN}} (\mathbf{\omega})=}
		\frac{\left| \left| \mathbf{\omega} \right| \right|^{2}_{1}}{\left| \left| \mathbf{\omega} \right| \right|^{2}_{2}}=\frac{\left(\sum_{i \in [n]} \omega_{i}^{2} \right)^2}{\sum_{i \in [n]} \omega_{i}^{2}}&\emph{if} $\mathbf{\omega}\neq0$. \nonumber\\
   		0&\emph{if }$\mathbf{\omega}=0$. \nonumber
	\end{numcases}
	While the channel model of Section~\ref{subsec:channelmodel} assumed a hard-decision on the AWGN-channel outputs,~\eqref{eq:pairwise}
	holds even when log-likelihood ratios (which have no loss in optimality) are used in message-passing~\cite{KoetterVontobel1}. Hence, 
	we can use~\eqref{eq:pairwise} to obtain a lower bound for \emph{any} message-passing decoder. The \emph{minimum} pseudoweight 
	$w_{p}^{\mathrm{AWGN,min}}$ of a parity-check matrix of a given code is defined as the minimum AWGN pseudoweight 
	over all \emph{nonzero} pseudo-codewords of the code~\cite[Definition 37]{KoetterVontobel1}. For any
	regular-LDPC code of blocklength $n$ with $d_v \geq 3$, the minimum AWGN pseudoweight is upper-bounded 
	as~\cite[Proposition 49]{KoetterVontobel1},~\cite[Theorem 7]{KoetterVontobel2}:
	\begin{equation}
		\label{eq:minpseudobound} w_{p}^{\mathrm{AWGN,min}} (\mathbf{\omega}) \leq \left( \frac{d_v (d_v - 1)}{(d_v - 2)} \right)^2 n^{\frac{2 \log (d_v - 1)}{\log (d_v - 1) (d_c - 1)}}.
	\end{equation}
	Therefore, lower bounding the word-error probability $P^{word}_{e}$ by the pairwise error-probability and using~\eqref{eq:minpseudobound} in~\eqref{eq:pairwise}:
	\begin{equation}
	P^{word}_{e} \geq P_{0 \to \mathbf{\omega}} \geq \mathbb{Q} \left( \sqrt{2 \frac{E_s}{N_0} \left( \frac{d_v (d_v - 1)}{(d_v - 2)} \right)^2 n^{\frac{2 \log (d_v - 1)}{\log (d_v - 1) (d_c - 1)}}} \right).
	\end{equation}
	 Using our notation from Section~\ref{subsec:channelmodel}, $\frac{E_s}{N_0} = \eta P_{T}$, and trivially bounding bit-error probability $P_e$~\cite[Eqn.~(2)]{blockvsbiterrors}:
	 \begin{eqnarray}
	  \nonumber && P_{e} \geq \frac{P^{word}_{e}}{n}\\
	  \nonumber &\geq& \frac{\mathbb{Q} \left( \sqrt{2 \eta P_{T} \left( \frac{d_v (d_v - 1)}{(d_v - 2)} \right)^2 n^{\frac{2 \log (d_v - 1)}{\log (d_v - 1) (d_c - 1)}}} \right)}{n}\\
	  \nonumber &\overset{\bullet}{>}& \frac{ \frac{1}{n} e^{-\eta P_{T} \left( \frac{d_v (d_v - 1)}{d_v - 2} \right)^2 n^{\frac{2 \log (d_v - 1)}{\log (d_v - 1) (d_c - 1)}}}}{\left(n^{\frac{1}{1 + \frac{\log (d_c - 1)}{\log (d_v - 1)}}}2 \sqrt{\pi \eta P_{T}} \left( \frac{d_v (d_v - 1)}{(d_v - 2)} \right) + \sqrt{\frac{\pi}{\eta P_{T}}}\frac{d_v - 2}{d_v (d_v - 1)} \right)}\\
	   \label{eq:lemma4step2} &\overset{(\ast)}{>}& \frac{e^{-\eta P_{T} \left( \frac{d_v (d_v - 1)}{d_v - 2} \right)^2 n^{\frac{2 \log (d_v - 1)}{\log (d_v - 1) (d_c - 1)}}}}{n^{1 +\frac{\log (d_v - 1)}{\log (d_v - 1) (d_c - 1)}} \left( \frac{d_v (d_v - 1)}{(d_v - 2)} \right) \left( 2 \sqrt{\pi \eta P_{T}} + \sqrt{\frac{\pi}{\eta P_{T}}} \right)}
	\end{eqnarray}
	where $(\bullet)$ holds because of~\eqref{eq:millsratioofficial} and $n \geq 1$, and  $(\ast)$ 
	holds because $n \geq 1$ and  $d_v (d_v -1) > (d_v - 2)$. It follows that whenever 
	$P_{T} \geq \frac{1}{\eta}$,
	 \begin{eqnarray}
	 \nonumber P_{e} &>& \frac{e^{-\eta P_{T} \left( \frac{d_v (d_v - 1)}{d_v - 2} \right)^2 n^{\frac{2 \log (d_v - 1)}{\log (d_v - 1) (d_c - 1)}}}}{n^{1 +\frac{\log (d_v - 1)}{\log (d_v - 1) (d_c - 1)}} \left( 3 \frac{d_v (d_v - 1)}{(d_v - 2)} \sqrt{\pi \eta P_{T}} \right)}\\
	 \label{eq:lemma4step4} &\overset{(\star)}{>}& \frac{e^{-\eta P_{T} (1 + 9 \pi) \left( \frac{d_v (d_v - 1)}{d_v - 2} \right)^2 n^{\frac{2 \log (d_v - 1)}{\log (d_v - 1) (d_c - 1)}}}}{n}
	\end{eqnarray}
	where $(\star)$ holds because $e^{-x^2} < \frac{1}{x}$ for all $x \geq 0$. Inverting both sides of~\eqref{eq:lemma4step4}, taking $\log (\cdot)$ and then simplifying,
	\begin{eqnarray}
	\nonumber &&\frac{\log \frac{1}{P_e}}{\eta P_{T} (1 + 9 \pi) \left(\frac{d_v (d_v - 1)}{(d_v - 2)} \right)^{2}} < n^\frac{2}{1 + \frac{\log(d_c - 1)}{\log (d_v - 1)}}\\
	 \nonumber  &+& \frac{ \left(\frac{(d_v - 2)}{d_v (d_v - 1)}\right)^{2} \log n}{\eta P_{T} (1 + 9 \pi)}\\
	 \label{eq:lemma4step7} &\overset{P_{T} \geq \frac{1}{\eta}}{<}& n^\frac{2}{1 + \frac{\log(d_c - 1)}{\log (d_v - 1)}} + \log n
	\end{eqnarray}
	We have shown that~\eqref{eq:lemma4step7} holds for any $P_{e}$ and any 
	$P_{T} \geq \frac{1}{\eta}$, hence ignoring the non-dominating term on the RHS and then 
	raising both sides to the power $\frac{\log(d_v - 1)(d_c - 1)}{2 \log (d_v - 1)}$, we get the desired result. For the case 
	when $d_v = 2$, because the minimum distance of regular-LDPC codes with $d_v = 2$ is at most $2 + \frac{2 \log \frac{n}{2}}{\log(d_c - 1)}$ 
	(see~\cite[Theorem 2.5]{gallagerthesis}), the pairwise error-probability that a minimum-weight nonzero \emph{codeword} $x'$ is 
	decoded when the all-zero codeword was transmitted is 
	\begin{equation}
		\label{eq:pairwise2} P_{0 \to x'} \geq \mathbb{Q} \left( \sqrt{2 \frac{E_s}{N_0}  \left(2 + \frac{2 \log \frac{n}{2}}{\log (d_c - 1)} \right)} \right),
	\end{equation}
	Replacing $\frac{E_s}{N_0}$ by $\eta P_T$, the word-error probability is
	\begin{equation}
		P^{word}_{e} \geq P_{0 \to x'}  = \mathbb{Q} \left( \sqrt{2 \eta P_{T} \left(2 + \frac{2 \log \frac{n}{2}}{\log (d_c - 1)} \right)} \right).
	\end{equation}
	Then applying an identical analysis to the $d_v > 3$ case, for any bit-error probability $P_{e}$:
	\begin{eqnarray}
	\nonumber P_{e} \geq \frac{\frac{1}{n} e^{-\eta P_{T} (2 + \frac{2 \log \frac{n}{2}}{\log (d_c - 1)})}}{\left(2 \sqrt{\pi \eta P_{T} \left(2 + \frac{2 \log \frac{n}{2}}{\log(d_c - 1)} \right)} + \sqrt{\frac{\pi}{\eta P_{T}\left(2 + \frac{2 \log \frac{n}{2}}{\log(d_c - 1)} \right)}} \right)}\\
	\nonumber \overset{n \geq 1}{>} \frac{\frac{1}{n} e^{-\eta P_{T} (2 + \frac{2 \log \frac{n}{2}}{\log (d_c - 1)} )}}{\left(2 \sqrt{\pi \eta P_{T} \left(2 + \frac{2 \log \frac{n}{2}}{\log(d_c - 1)} \right)} + \sqrt{\frac{\pi \left(2 + \frac{2 \log \frac{n}{2}}{\log(d_c - 1)} \right)}{\eta P_{T}}} \right)}.
	\end{eqnarray}
	Then for any $P_{T} \geq \frac{1}{\eta}$, we also have
	\begin{equation}
	 \label{eq:lemma4step10} P_{e} > \frac{e^{-\eta P_{T} (2 + \frac{2 \log \frac{n}{2}}{\log (d_c - 1)})}}{3n \sqrt{\pi \eta P_{T} \left(2 + \frac{2 \log \frac{n}{2}}{\log(d_c - 1)} \right)}} \overset{(\star)}{>} \frac{e^{-\eta P_{T} (1 + 9 \pi) (2 + \frac{2 \log \frac{n}{2}}{\log (d_c - 1)} )}}{n},
	\end{equation}
	where $(\star)$ holds because $e^{-x^2} < \frac{1}{x}$ for all $x \geq 0$. Inverting both sides 
	of~\eqref{eq:lemma4step10}, taking $\log (\cdot)$ and then simplifying,
	\begin{eqnarray}
	\nonumber \frac{\log \frac{1}{P_e}}{\eta P_{T} (1 + 9 \pi)} < 2 + \frac{2 \log \frac{n}{2}}{\log (d_c - 1)} + \frac{\log n}{\eta P_{T}  (1 + 9 \pi)}\\
	\nonumber \frac{\log \frac{1}{P_e}}{\eta P_{T} (1 + 9 \pi)} \overset{P_{T} \geq \frac{1}{\eta}}{<}  2 + \frac{2 \log n - 2 \log 2}{\log (d_c - 1)}  + \log n\\
	\nonumber \overset{n \geq 1}{\leq}  2 + \frac{2 \log n - 2 \log 2}{\log (d_c - 1)} + 2 \log n\\
	\label{eq:lemma4step11} < \left(2 + \frac{2}{\log (d_c - 1)} \right) \left(1 + \log n \right). 
	\end{eqnarray}
	Dividing both sides of~\eqref{eq:lemma4step11} by $ \left(2 + \frac{2}{\log (d_c - 1)} \right)$, taking $e^{( \cdot)}$ on both sides, and simplifying:
	\begin{equation}
	n > \frac{1}{e} \left( \frac{1}{P_e} \right)^{\frac{1}{\eta P_{T} (1 + 9 \pi) \left( 2 + \frac{2}{\log (d_c - 1)} \right)}},
	\end{equation}
	which completes the proof of Lemma~\ref{lemma:generalblocklength}.
	\end{proofof}

	\section{Proof of Lemma~\ref{lem:itera}}
	\allowdisplaybreaks
	\label{app:proofitera}
	\begin{proofof}{Lemma~\ref{lem:itera}}
	First, note the $\Theta \left( \cdot \right)$ expression in Lemma~\ref{lem:itera} 
	contains two variables: $\frac{1}{P_e}$ and $P_{T}$. We analyze
	the minimum number of independent iterations as a function 
	$N_{\mathrm{iter}}: [2,\infty) \times [\Delta_A,\infty) \to \mathbb{R}^{\geq 0}$ 
	(Definition~\ref{def:bigonotationmult}), where $\Delta_{A} > 0$ is the transmit power for which
	$p_0$ is exactly the \emph{threshold} for decoding over the BSC~\cite[Section 4.3]{gallagerthesis},~\cite{urbankecapacity}. 
	Explicitly, if $\sigma_A$ is the threshold for Gallager-A decoding over the BSC, $\mathbb{Q} \left( \sqrt{2 \eta \Delta_{A}} \right) = \sigma_A$.
	
	Note when $P_T < \Delta_{A}$, it is not possible to force $P_e \to 0$, hence $N_{\mathrm{iter}}$ 
	will be infinite for all $P_{e}$ below some constant~\cite{urbankecapacity}. No further analysis 
	is needed for such low transmit powers, since all $P_{e}$ above said constant 
	can be achieved with $\Theta \left( 1\right)$ transmit power and $\mathcal{O} \left(1 \right)$ 
	decoding iterations. From~\cite[Eqn.~(6)]{urbankecapacity}, the bit-error probability 
	after the $i$th decoding iteration, $p_i$, is
	\begin{eqnarray}
	\nonumber p_{i} &=& p_{0} - p_{0} \left[\frac{1 + (1-2p_{i-1})^{dc-1}}{2} \right]^{d_v - 1}\\
	 \label{eq:recurrenceA}  &+& (1-p_{0})\left[ \frac{1 - (1-2p_{i-1})^{dc-1}}{2} \right]^{d_v - 1}.
	\end{eqnarray}
	Since the RHS of~\eqref{eq:recurrenceA} is differentiable with respect to (w.r.t.) $p_{i-1}$, 
	by Taylor's Theorem there exists a real function $R_1(x)$ 
	with $\lim_{x \to 0} R_1(x) = 0$ such that:
	\begin{equation}
	\label{eq:taylorA} p_{i} = p_0 (d_v - 1)(d_c - 1) p_{i-1} + R_1(p_{i-1}).
	\end{equation}
	The RHS of~\eqref{eq:taylorA} is the first-order MacLaurin expansion of $p_{i}$. 
	Further, because the RHS of~\eqref{eq:recurrenceA} is a polynomial in $p_{i-1}$, it is 
	twice continuously differentiable and by the mean value theorem the remainder term $R_1(p_{i-1})$ has Lagrange form:
	\begin{equation}
	\label{eq:lagrangetermA} R_1(p_{i-1}) = \frac{1}{2} \frac{d^2 p_{i}(x^{*})}{d p_{i-1} ^2}p_{i-1}^2,
	\end{equation}
	where $x^{*} \in (0, p_{i-1})$. It can be verified that the second derivative of $p_i$ w.r.t. $p_{i-1}$
%	we find that it
%	\begin{align}
%	\label{eq:secondderivA} \frac{d^2 p_{i}}{d p_{i-1}^2} &=-\frac{p_0}{2^{dv-3}}(d_v-1)(d_v-2)(d_c-1)^2 \left[1+(1-2p_{i-1})^{d_c-1}\right]^{d_v-3}(1-2p_{i-1})^{2(d_c-2)}\\
%	\nonumber &-\frac{p_0}{2^{dv-3}}(d_v - 1)(d_c - 1)(d_c - 2)\left[1+(1-2p_{i-1})^{d_c-1}\right]^{d_v-2}(1-2p_{i-1})^{d_c-3}\\
%	\nonumber &+\frac{(1-p_0)}{2^{dv-3}}(d_v - 1)(d_v - 2)(d_c-1)^2 \left[1-(1-2p_{i-1})^{d_c-1}\right]^{d_v-3}(1-2p_{i-1})^{2(d_c-2)}\\
%	\nonumber &-\frac{(1-p_0)}{2^{dv-3}}(d_v - 1)(d_c - 1)(d_c - 2)\left[1-(1-2p_{i-1})^{d_c-1}\right]^{d_v-2}(1-2p_{i-1})^{d_c-3}.
%	\end{align}
%	It can be verified that the derivative of the RHS of~\eqref{eq:secondderivA} w.r.t. $p_{i-1}$ 
%	is nonnegative for $p_{i-1} \in \left[0,\frac{1}{2}\right]$. Therefore, the RHS of~\eqref{eq:secondderivA} 
	is minimized at $p_{i-1} = 0$ and maximized at $p_{i-1} = \frac{1}{2}$. Solving for 
	both cases and plugging into~\eqref{eq:lagrangetermA}, we find
	\begin{align}
		\nonumber -p_0 (d_v - 1)(d_c - 1) &\left[ \frac{(d_v - 2)(d_c - 1)}{2} + (d_c - 2) \right]p_{i-1}^2\\
		\label{eq:R1upperboundA} &\leq R_1 \leq 0
	\end{align}
	Plugging~\eqref{eq:R1upperboundA} into~\eqref{eq:taylorA} and applying the RHS recursively, the 
	bit-error probability after $i$th decoding iteration $p_i$ is:
	\begin{align}
		\nonumber &p_{0}(d_v - 1)(d_c - 1)p_{i-1} \Bigg[1 - p_{i-1} \Bigg( \frac{(d_v - 2)(d_c - 1)}{2}\\
		\label{eq:upperboundpiA}  &+ (d_c - 2) \Bigg) \Bigg] \leq p_{i} \leq \left[p_{0}(d_v - 1)(d_c - 1)\right]^{i}.
	\end{align}
	 Now, choose an arbitrary $0 < \delta < \frac{1}{2}$ and choose $P_T$ (thereby $p_0$ as well). As explained in~\cite[Section 4.3]{gallagerthesis}, 
	 since we are operating above the threshold, we are guaranteed that $p_{i} < p_{i-1} \leq p_0$ and $p_i \overset{i \to \infty}{\rightarrow} 0$. Thus, 
	 for sufficiently small $p_{i}$ (thereby small $p_{i-1}$) \emph{or} sufficiently large $P_T$ (thereby small $p_{0}$),~\eqref{eq:upperboundpiA} becomes:
	\begin{equation}
		\label{eq:smallerrorprobA} p_{0}(d_v - 1)(d_c - 1)(1 - \delta) p_{i-1} \leq p_{i} \leq \left[ p_{0}(d_v - 1)(d_c - 1) \right]^{i}
	\end{equation}
	Applying the relation on the LHS of~\eqref{eq:smallerrorprobA} recursively 
	\begin{align}
		\nonumber & p_0 \left[\frac{p_0}{2}(d_v - 1)(d_c - 1) \right]^{i} \overset{\delta < \frac{1}{2}}{\leq} p_{i} \leq \left[p_{0}(d_v - 1)(d_c - 1)\right]^{i}\\
		\nonumber &\frac{ \sqrt{\frac{\eta}{\pi} P_T} e^{-\eta P_T}}{\left(2 \eta P_T + 1\right)} \left[\frac{(d_v - 1)(d_c - 1)\sqrt{\frac{\eta}{\pi} P_T} e^{-\eta P_T}}{2 \left(2 \eta P_T + 1\right)}\right]^{i} \overset{\eqref{eq:millsratioofficial}}{\leq} p_{i}\\
		\label{eq:millsboundssmallpA} & \overset{\eqref{eq:millsratioofficial}}{\leq} \left[(d_v - 1)(d_c - 1)\frac{e^{-\eta P_T}}{\sqrt{4 \pi \eta P_T}}\right]^{i}.
	\end{align} 
	Inverting all sides of~\eqref{eq:millsboundssmallpA}, 
	taking $\log(\cdot)$ on all sides, replacing $p_i$ by $P_e$ and $i$ 
	by $N_{\mathrm{iter}}$, and dividing all sides by $\eta P_T$:
	\begin{align}
	\nonumber &N_{\mathrm{iter}} \left[1 + \frac{\log \eta P_T }{2 \eta P_T} - \frac{\log (d_v - 1)(d_c -1)}{\eta P_T} + \frac{\log 2 \sqrt{\pi}}{\eta P_T} \right]\\
	\nonumber &\leq \frac{\log \frac{1}{P_e}}{\eta P_T} \leq N_{\mathrm{iter}} \bigg[ 1- \frac{\log \frac{\eta}{\pi} P_T}{2 \eta P_T} - \frac{\log 0.5 (d_v - 1)(d_c - 1)}{\eta P_T}\\
	\label{eq:sincefixed} &+ \frac{\log \left( 2 \eta P_T + 1 \right)}{\eta P_T} \bigg] + 1 + \frac{\log \left( 2 \eta P_T + 1 \right)}{\eta P_T} - \frac{\log \frac{\eta}{\pi} P_T}{2 \eta P_T}. 
	\end{align}
	We have shown that~\eqref{eq:sincefixed} holds for any choice of $P_{T}$ as long as 
	$P_{e}$ is sufficiently small, which completes the proof of the constant $P_{T}$ result. 
	Next, set $P_{T} \geq \max \{\frac{2 \log (d_v - 1) (d_c - 1)}{\eta}, \frac{\pi}{\eta} \}$. As 
	explained above,~\eqref{eq:sincefixed} also holds for any $P_e$ as long as $P_T$ is sufficiently large. 
	In this case~\eqref{eq:sincefixed} simplifies to 
	\begin{align}
	\nonumber &N_{\mathrm{iter}} \left[1 - \frac{\log (d_v - 1)(d_c -1)}{\eta P_T} \right] \overset{P_T > \frac{1}{\eta} > 0}{\leq} \frac{\log \frac{1}{P_e}}{\eta P_T}\\
	\nonumber &\overset{P_T \geq \frac{\pi}{\eta} > \frac{1}{\eta}; d_c > d_v \geq 2}{\leq} N_{\mathrm{iter}} \left[1 +  \log 3 \right] + 1 + \log 3\\
	&\frac{1}{2} N_{\mathrm{iter}} \overset{P_T \geq \frac{2 \log (d_v - 1) (d_c - 1)}{\eta}}{\leq} \frac{\log \frac{1}{P_e}}{\eta P_T} < \left[1+ \log 3  \right] N_{\mathrm{iter}} + 3,
	\end{align}
	which completes the proof of the Lemma.
	\end{proofof}
	
	\section{Proof of Lemma~\ref{lem:iterb}}
	\allowdisplaybreaks
	\label{app:proofiterb}
	\begin{proofof}{Lemma~\ref{lem:iterb}}
	We analyze the number of independent iterations as a 
	function $N_{\mathrm{iter}}: [2,\infty) \to \mathbb{R}^{\geq 0}$, since even in the second case,
	the transmit power is a function of $\frac{1}{P_e}$. Let $\Delta_{B} > 0$ be 
	the transmit power for which $p_0$ is exactly the \emph{threshold} for decoding over the 
	BSC~\cite[Section 4.3]{gallagerthesis},~\cite{urbankecapacity}. As explained in the proof of Lemma~\ref{lem:itera}, we 
	need not consider cases where $P_T < \Delta_{B}$. Using~\cite[Eqn. 4.15]{gallagerthesis}, 
	for $d_v$ odd, the bit-error probability after the $i$th decoding iteration follows
	\begin{align}
	\nonumber p_{i} &= p_0 - \frac{p_0}{2^{dv-1}} \sum_{m = \frac{d_v-1}{2}}^{d_v-1}\binom{d_v-1}{m} \left[1 + (1-2p_{i-1})^{d_c-1} \right]^{m}\\
	\nonumber &\times \left[1 - (1-2p_{i-1})^{d_c-1} \right]^{d_v-1-m}\\
	\nonumber &+ \frac{1-p_0}{2^{dv-1}}\sum_{m = \frac{d_v-1}{2}}^{d_v-1}\binom{d_v-1}{m} \left[1 - (1-2p_{i-1})^{d_c-1} \right]^{m}\\
	\label{eq:gal1} &\times \left[1 + (1-2p_{i-1})^{d_c-1} \right]^{d_v-1-m}.
	\end{align}
	The RHS of~\eqref{eq:gal1} is a polynomial in $p_{i-1}$ and the $\frac{d_v-1}{2}$th order Maclaurin expansion is 
	\begin{equation}
	\label{eq:taylorB} p_{i} = p_0 \binom{d_v-1}{\frac{d_v - 1}{2}} \left(d_c - 1\right)^{\frac{d_v - 1}{2}}p_{i-1}^{\frac{d_v - 1}{2}} + R_B(p_{i-1}).
	\end{equation}
	Because the RHS of~\eqref{eq:gal1} is a polynomial, by the mean value theorem, the remainder has a Lagrange form (where $x^{*} \in (0, p_{i-1})$):
	\begin{equation}
	\label{eq:lagrangetermB} R_B(p_{i-1}) = \frac{1}{\left(\frac{d_v + 1}{2}\right)!} \frac{d^{\frac{d_v + 1}{2}} p_{i} (x^{*})}{d p_{i-1} ^{\frac{d_v + 1}{2}}} p_{i-1}^{\frac{d_v + 1}{2}}.
	\end{equation}
	The $\frac{d_v + 1}{2}$th derivative of $p_{i}$ is another polynomial; therefore it must be bounded on the bounded interval $[0,\frac{1}{2}]$ and
	\begin{equation}
	\label{eq:lagrangetermB2} - c_{l}^{B} p_{i-1}^{\frac{d_v + 1}{2}} \leq R_B(p_{i-1}) \leq c_{u}^{B} p_{i-1}^{\frac{d_v + 1}{2}},
	\end{equation}	 
	for some constants $c_{l}^{B}, c_{u}^{B} > 0$. Then choose $P_T$. 
	Since we exceed the decoding threshold~\cite{urbankecapacity}, $p_i < p_{i-1} \leq p_0$. Now, take $i$ to be the final iteration. 
	Since we assumed $\lim_{P_e \to 0} \frac{P_T}{\log \frac{1}{P_e}} = 0$, we will also have $\lim_{P_e \to 0} \frac{p_{i-1}}{p_0} = 0$ 
	(the number of decoding iterations used in the coding strategy eventually exceeds $1$ as $P_e \to 0$). Hence, for sufficiently 
	small $p_{i}$ (thereby small $p_{i-1}$), 
	\begin{align}
	\nonumber &-\frac{1}{2} p_0 \binom{d_v-1}{\frac{d_v - 1}{2}} \left(d_c - 1\right)^{\frac{d_v - 1}{2}} p_{i-1}^{\frac{d_v - 1}{2}} \leq R_B(p_{i-1})\\
	\label{eq:lagrangetermBbd}  &\leq \frac{1}{2} p_0 \binom{d_v-1}{\frac{d_v - 1}{2}} \left(d_c - 1\right)^{\frac{d_v - 1}{2}} p_{i-1}^{\frac{d_v - 1}{2}}.
	\end{align}
	Plugging~\eqref{eq:lagrangetermBbd} into~\eqref{eq:taylorB}, we have
	\begin{align}
	 \nonumber &p_0 \left[ \binom{d_v-1}{\frac{d_v - 1}{2}} \left(d_c - 1\right)^{\frac{d_v - 1}{2}} \frac{1}{2} \right] p_{i-1}^{\frac{d_v - 1}{2}} \leq p_{i}\\
	 \label{eq:galbbigtheta} &\leq p_0  \left[ \binom{d_v-1}{\frac{d_v - 1}{2}} \left(d_c - 1\right)^{\frac{d_v - 1}{2}} \frac{3}{2} \right] p_{i-1}^{\frac{d_v - 1}{2}}.
	\end{align}
	Applying~\eqref{eq:galbbigtheta} recursively, we obtain
	\begin{align}
	\nonumber &\bigg[ p_{0}^{1 + \cdots + \left(\frac{d_v - 1}{2} \right)^{i}} \left( \binom{d_v-1}{\frac{d_v-1}{2}} \frac{1}{2} \right)^{1 + \cdots + \left(\frac{d_v - 1}{2} \right)^{i-1}}\\
	\nonumber &\times (d_c - 1)^{\left(\frac{d_v - 1}{2} \right) + \cdots + \left(\frac{d_v - 1}{2} \right)^{i}} \bigg] \leq p_{i}\\
	\nonumber &\leq \bigg[p_{0}^{1 + \cdots + \left(\frac{d_v - 1}{2} \right)^{i}} \left( \binom{d_v-1}{\frac{d_v-1}{2}} \frac{3}{2} \right)^{1 + \cdots + \left(\frac{d_v - 1}{2} \right)^{i-1}}\\
	\label{eq:galbbigthetasimple} &\times (d_c - 1)^{\left(\frac{d_v - 1}{2} \right) + \cdots + \left(\frac{d_v - 1}{2} \right)^{i}} \bigg].
	\end{align}
	Loosening the LHS and RHS of~\eqref{eq:galbbigthetasimple} and grouping like-terms we have
	\begin{align}
	\nonumber &\left[ p_{0}^{1 + \cdots + \left(\frac{d_v - 1}{2} \right)^{i}} \left( \binom{d_v-1}{\frac{d_v-1}{2}} (d_c-1) \frac{1}{2} \right)^{1 + \cdots + \left(\frac{d_v - 1}{2} \right)^{i-1}} \right]\\
	\nonumber &\overset{(i \geq 0; d_c > 2)}{\leq} p_{i} \overset{(p_0 \leq 1; d_c > 2)}{\leq} \left(p_{0} \binom{d_v-1}{\frac{d_v-1}{2}} \frac{3}{2} \right)^{1 + \cdots + \left(\frac{d_v - 1}{2} \right)^{i-1}}\\ 
	\label{eq:galbbigthetasimple2} &\times (d_c - 1)^{1 + \cdots +\left(\frac{d_v - 1}{2} \right)^{i}}.
	\end{align}
	Simplifying geometric progressions, inverting 
	all sides of~\eqref{eq:galbbigthetasimple2}, taking $\log(\cdot)$ on all sides, and replacing 
	$p_i$ by $P_e$ and $i$ by $N_{\mathrm{iter}}$:
	\begin{align}
	\nonumber &\frac{\left( \frac{d_v - 1}{2} \right)^{N_{\mathrm{iter}}} - 1}{\left( \frac{d_v - 1}{2} \right) - 1} \left[ \log \frac{1}{p_0} + \log \frac{2}{3\binom{d_v-1}{\frac{d_v-1}{2}}(d_c - 1)} \right] \leq \log \frac{1}{P_e}\\ 
	\nonumber &\leq \frac{\left( \frac{d_v - 1}{2} \right)^{N_{\mathrm{iter}} + 1} - 1}{\left( \frac{d_v - 1}{2} \right) - 1} \log \frac{1}{p_0}\\ 
	\label{eq:galbgeo} &+ \frac{\left( \frac{d_v - 1}{2} \right)^{N_{\mathrm{iter}}} - 1}{\left( \frac{d_v - 1}{2} \right) - 1}\log \frac{2}{\binom{d_v-1}{\frac{d_v-1}{2}} (d_c - 1)}.
	\end{align}
	Applying~\eqref{eq:millsratioofficial} on $p_0$ terms in~\eqref{eq:galbgeo}, dividing all sides by $\eta P_{T}$, 
	loosening the RHS by ignoring negative terms, and simplifying:
	\begin{align}
	 \nonumber &\frac{\left( \frac{d_v - 1}{2} \right)^{N_{\mathrm{iter}}} - 1}{\left( d_v - 3 \right)} \left[ 2 + \frac{\log 4 \pi \eta P_{T} -2 \log \frac{3}{2} \binom{d_v-1}{\frac{d_v-1}{2}} (d_c - 1)}{\eta P_{T}} \right]\\
	 \nonumber &\leq \frac{\log \frac{1}{P_e}}{\eta P_{T}} \leq \frac{\log (d_c - 1)}{\eta P_T}\\ 
	 \label{eq:galbgeo2} &+ \frac{\left( \frac{d_v - 1}{2} \right)^{N_{\mathrm{iter}} + 1}}{\left( d_v - 3 \right)} \left[2 + \frac{2 \log \left( \sqrt{4 \pi \eta P_{T}} + \sqrt{\frac{\pi}{\eta P_{T}}} \right)}{\eta P_{T}} \right].
	\end{align}
	We have shown that~\eqref{eq:galbgeo2} holds for any $P_{T}$ as long as $P_{e}$ is sufficiently small. Thus, treating $P_{T}$ 
	as a constant in~\eqref{eq:galbgeo2} and taking $\log ( \cdot )$ on all sides completes the proof of the fixed $P_{T}$ result. 
	For the other case, consider the limits of the leftmost and rightmost side of~\eqref{eq:galbgeo2} as $P_{T} \to \infty$.
	For any $\epsilon > 0$, for sufficiently large $P_{T}$ the following holds:
	\begin{align}
	\nonumber &\left(\left( \frac{d_v - 1}{2} \right)^{N_{\mathrm{iter}}} - 1\right) \frac{2 - \epsilon}{\left( d_v - 3 \right)} \leq \frac{\log \frac{1}{P_e}}{\eta P_{T}}\\
	\label{eq:galbgeo3} &\leq \left( \frac{d_v - 1}{2} \right)^{N_{\mathrm{iter}} + 1} \frac{2 + \epsilon}{\left( d_v - 3 \right)}.
	\end{align}	
	Taking $\log(\cdot)$ on all sides of~\eqref{eq:galbgeo3} and simplifying, we obtain:
	\begin{align}
	\nonumber &\log \left( \left( \frac{d_v - 1}{2} \right)^{N_{\mathrm{iter}}} - 1\right) + \log \frac{2 - \epsilon}{\left( d_v - 3 \right)} \leq \log \frac{\log \frac{1}{P_e}}{\eta P_{T}}\\
	\label{eq:galbgeolast} &\leq \left(N_{\mathrm{iter}} + 1\right) \log \left( \frac{d_v - 1}{2}\right) + \log \frac{2 + \epsilon}{\left( d_v - 3 \right)},
	\end{align}	
	which is equivalent to the desired result.
	\end{proofof}
	
	\vspace{-0.2cm}
	\section{Proof of Theorem~\ref{thm:anyLDPClower}}
	\label{app:proofthmany}	
	\allowdisplaybreaks
	\begin{proofof}{Theorem~\ref{thm:anyLDPClower}}
	Since we are proving a lower bound, we can restrict ourselves to the case where $d_v \geq 3$ 
	without loss of generality (the decoding power when $d_v = 2$ grows exponentially faster).
	Via Lemma~\ref{lemma:generalblocklength} and Lemma~\ref{lem:triviallower}, the
	total power required for a $(d_v, d_c)$-regular LDPC code and any iterative message-passing decoder to achieve
	bit-error probability $P_e$ under the Wire Model is
	\begin{equation}
	\label{eq:step1thm4} P_{\mathrm{total}} = \Omega \left( P_{T} + \left(\frac{\log \frac{1}{P_e}}{\eta P_{T} (1 + 9 \pi) \left(\frac{d_v (d_v - 1)}{d_v - 2} \right)^{2} } \right)^{\frac{1 + \frac{\log(d_c - 1)}{\log(d_v - 1)}}{2}}\right)
	\end{equation}
	
	First, it follows from~\eqref{eq:lemma4step2} that if $P_T$ is kept fixed while $P_e \to 0$, the total 
	power (and the decoding power) diverges as
	\begin{equation}
		\label{eq:step2thm4} P_{\mathrm{total,bdd\;P_T}}=\Omega\left(\log^{\frac{1 + \frac{\log (d_c - 1)}{\log (d_v - 1)}}{2}}\frac{1}{P_e}\right).
	\end{equation}
	The exponent of $\log \frac{1}{P_e}$ in~\eqref{eq:step2thm4} is always greater than $1$ since $d_c > d_v$ for any regular-LDPC code. 
	Next, differentiating the expressions inside the $\Omega \left( \cdot \right)$ of~\eqref{eq:step1thm4} 
	w.r.t. $P_T$, setting to zero, and substituting the minimizing transmit power into~\eqref{eq:step1thm4} we 
	find that the minimum total power is:
	\begin{equation}
		\label{eq:step3thm4} P_{\mathrm{total,min}}=\Omega\left(\log^{\frac{1}{1 + \frac{2}{1 + \frac{\log (d_c - 1)}{\log (d_v - 1)} }}} \frac{1}{P_e}\right),
	\end{equation}
	which completes the proof of the theorem.
	\end{proofof}
	
	\section{Proof of Theorem~\ref{thm:gallageratotalpower}}
	\label{app:proofthma}
	\allowdisplaybreaks
	\begin{proofof}{Theorem~\ref{thm:gallageratotalpower}}
	Let $N^{(P_e)}_{\mathrm{iter}}$ denote the minimum number of independent 
	Gallager-A decoding iterations required to achieve bit-error probability $P_e$. 
	Via Theorem~\ref{thm:asymlbwire}, the total power is lower bounded by $P_{\mathrm{total}} = P_T + \Omega \left(e^{\gamma N^{(P_e)}_{\mathrm{iter}}} \right)$. 
	It follows from Lemma~\ref{lem:itera} that if $P_T$ is kept fixed as 
	$P_e \to 0$, then the required decoding power diverges atleast as 
	fast as a power of $\frac{1}{P_e}$, which is exponentially larger than 
	the power required for uncoded transmission. If instead the transmit power is 
	allowed to vary, it follows from Lemma~\ref{lem:itera} that
	\begin{eqnarray}
		\label{eq:lowerboundexp}	P_{\mathrm{total}} &=& \Omega\left(P_T + \left( \frac{1}{P_e} \right)^{\frac{\gamma}{\eta P_T}}\right).\\
		\label{eq:upperboundexp} P_{\mathrm{total}} &=& \mathcal{O} \left(P_T + \left( \frac{1}{P_e} \right)^{\frac{2 \gamma}{\eta P_T}}\right)
	\end{eqnarray}
	In order to find the optimizing transmit power, let $L_{P_e}(P_T)$ denote the function in the $\Omega \left( \cdot \right)$ expression 
	of~\eqref{eq:lowerboundexp} and let $U_{P_e}(P_T)$ denote the 
	function in the $\mathcal{O} \left( \cdot \right)$ expression of~\eqref{eq:upperboundexp}:
	\begin{eqnarray}
		\label{eq:lpept} L_{P_e}(P_T) &=& P_T + \left( \frac{1}{P_e} \right)^{\frac{\gamma}{\eta P_T}}\\
		\label{eq:upept} U_{P_e}(P_T) &=& P_T + \left( \frac{1}{P_e} \right)^{\frac{2\gamma}{\eta P_T}}.
	\end{eqnarray}
	We start by analyzing the lower bound. To find the $P_T$ which minimizes $L_{P_e}$, we differentiate $L_{P_e}$ and set it to 0
	\begin{eqnarray}
		\nonumber\frac{d L_{P_e}}{d P_T} &=& 1 - e^{\frac{\gamma \log\frac{1}{P_e}}{\eta P_T}}\frac{\gamma \log\frac{1}{P_e}}{\eta P_T^2} = 0\\
		\label{eq:lhs} \Rightarrow \frac{P_T^2}{\frac{\gamma\log\frac{1}{P_e}}{\eta}} &=& e^{\frac{\gamma \log\frac{1}{P_e}}{\eta P_T}}.
	\end{eqnarray}
	Now, let $\mathcal{P} = \frac{P_T}{\sqrt{\frac{\gamma\log\frac{1}{P_e}}{\eta}}}$. Substituting into~\eqref{eq:lhs}, we get
	\begin{eqnarray}
		\mathcal{P}^{2} &=& e^{\frac{\sqrt{\frac{\gamma\log\frac{1}{P_e}}{\eta}}}{\mathcal{P}}} \Rightarrow 2 \mathcal{P} \log \mathcal{P} = \sqrt{\frac{\gamma\log\frac{1}{P_e}}{\eta}}\\
		\label{eq:lambertnext} &\Rightarrow& \log \mathcal{P} e^{\log \mathcal{P}} = \frac{1}{2} \sqrt{\frac{\gamma\log\frac{1}{P_e}}{\eta}}.
	\end{eqnarray}
	The positive, real valued solution to~\eqref{eq:lambertnext} is given by the principal branch $W_{0} ( \cdot )$ of the Lambert W function~\cite{KnuthLambert}. Explicitly, when $x, z \in \mathbb{R}^{\geq 0}$ satisfy the relation $x = z e^{z}$, we say $z = W_{0}(x)$. Hence we can write
	\begin{eqnarray}
		\label{eq:lambertzzz} \log \mathcal{P} &=& W_{0} \left( \frac{1}{2} \sqrt{\frac{\gamma\log\frac{1}{P_e}}{\eta}} \right)\\
		\Rightarrow \mathcal{P}  &=&  \frac{\log \mathcal{P} e^{\log \mathcal{P}}}{\log \mathcal{P}} \overset{\eqref{eq:lambertnext}\eqref{eq:lambertzzz}}{=} \frac{\frac{1}{2} \sqrt{\frac{\gamma\log\frac{1}{P_e}}{\eta}}}{W_{0} \left( \frac{1}{2} \sqrt{\frac{\gamma\log\frac{1}{P_e}}{\eta}} \right)}.
	\end{eqnarray}
	Rewriting $\mathcal{P}$ in terms of $P_T$ we find the optimizing transmit power
	\begin{eqnarray}
		\label{eq:lambertlimit} P_T^* &=& \frac{\frac{\gamma\log\frac{1}{P_e}}{\eta}}{2 W_{0} \left(\frac{1}{2} \sqrt{\frac{\gamma\log\frac{1}{P_e}}{\eta}} \right)}.
	\end{eqnarray}
	The first two terms in the asymptotic expansion of $W_{0} (x)$ as $x \to \infty$ are $\log(x) - \log \log (x)$~\cite{KnuthLambert}. In fact, $\forall x \geq e$~\cite{LambertBounds}:
	\begin{equation}
		\label{eq:lambertbound} \log(x) - \log \log (x) \leq W_{0} (x) \leq \log(x) - \frac{1}{2} \log \log (x).
	\end{equation}
	Using~\eqref{eq:lambertbound} in~\eqref{eq:lambertlimit}, and ignoring constant terms 
	in the resulting denominator, the optimizing transmit power is
	\begin{equation*}
		P_T^* = \Theta \left( \frac{\frac{\gamma}{\eta}\log\frac{1}{P_e}}{\log \log \frac{1}{P_e} - 2 \log \log \log^{\frac{1}{2}} \frac{1}{P_e}} \right).
	\end{equation*}
	Plugging back into $L_{P_e}$ in~\eqref{eq:lpept}, and ignoring non-dominating terms,
	we get the lower bound. An identical analysis of $U_{P_e}$ in~\eqref{eq:upept} gives the upper bound 
	and completes the proof.
%	\begin{equation*}
%		P_T^* = \mathcal{O} \left( \frac{\frac{2 \gamma}{\eta}\log\frac{1}{P_e}}{2 \log \frac{1}{2}\sqrt{\frac{2 \gamma}{\eta}} + \log \log \frac{1}{P_e} - \log \log \frac{1}{2} \sqrt{\frac{2 \gamma}{\eta}} - \log \log \log^{\frac{1}{2}} \frac{1}{P_e}} \right).
%	\end{equation*}
%	Plugging back into $U_{P_e}$ in~\eqref{eq:upept} we obtain
%	\begin{equation}
%		\label{eq:ignoredenomupper} P_{\mathrm{total,min}} = \mathcal{O} \left( \frac{\frac{2\gamma}{\eta}\log\frac{1}{P_e}}{2 \log \frac{1}{2}\sqrt{\frac{2\gamma}{\eta}} + \log \log \frac{1}{P_e} - \log \log \frac{1}{2} \sqrt{\frac{2\gamma}{\eta}} - \log \log \log^{\frac{1}{2}} \frac{1}{P_e}} + \frac{\frac{\gamma}{2 \eta}\log\frac{1}{P_e}}{\log \frac{1}{2} \sqrt{\frac{2\gamma}{\eta}} + \frac{1}{2} \log \log \frac{1}{P_e}} \right).
%	\end{equation}
%	Ignoring constants and non-dominating terms in the denominators of both the transmit and decoding power in~\eqref{eq:ignoredenomupper}, we get 
	\end{proofof}
	
	\section{Proof of Theorem~\ref{thm:gallagebatotalpowerlower}}
	\label{app:proofthmb}
	\allowdisplaybreaks	
	\begin{proofof}{Theorem~\ref{thm:gallagebatotalpowerlower}}
	Via Theorem~\ref{thm:asymlbwire} and Lemma~\ref{lem:iterb},
	if the transmit power is kept fixed as $P_e \to 0$, the total power is
	\begin{equation}
		\label{eq:galbfixedtotalw} P_{\mathrm{total,bdd\;P_T}}=\Omega \left( P_T + \log^{\gamma} \frac{1}{P_e} \right).
	\end{equation}
	If instead $P_T$ is allowed to scale as a function of $P_e$,
	\begin{eqnarray}
		\label{eq:galbsublogtotalw} P_{\mathrm{total}} &\overset{Theorem~\ref{thm:asymlbwire}}{=}& \Omega \left(P_T + e^{\gamma N^{(P_e)}_{\mathrm{iter}}}\right)\\
		&\overset{Lemma~\ref{lem:iterb}}{=}& \Omega\left(P_T + e^{\frac{\gamma}{\log\left(\frac{d_v-1}{2}\right)} \log\frac{\log\frac{1}{P_e}}{P_T}}\right)\\
		\label{eq:galbtotal} &=&  \Omega\left(P_T + \left(\frac{\log\frac{1}{P_e}}{P_T}\right)^{\frac{\gamma}{\log\left(\frac{d_v-1}{2}\right)}}\right).
	\end{eqnarray}
	Using the upper bound from Theorem~\ref{thm:asymlbwire}, 
	\begin{eqnarray}
		\label{eq:galbtotalup}P_{\mathrm{total}} &=& \mathcal{O}\left( P_T + \left(\frac{\log\frac{1}{P_e}}{P_T}\right)^{\frac{2 \gamma}{\log\left(\frac{d_v-1}{2}\right)}}\right).
	\end{eqnarray}
	Then, considering the bounds on $\gamma$ in Theorem~\ref{thm:asymlbwire}, we examine 
	the exponents of $\log \frac{1}{P_e}$ in~\eqref{eq:galbfixedtotalw} and $\frac{\log\frac{1}{P_e}}{P_T}$ in~\eqref{eq:galbtotal}:
	\begin{align}
		\nonumber \gamma &\geq \log{(d_v - 1)} + \log{(d_c - 1)} \overset{d_c > d_v > 3}{\geq}  \log 12\\
& \label{eq:ratio}\Rightarrow\frac{\gamma}{\log\left(\frac{d_v-1}{2}\right)} > \frac{1 + \frac{\log{(d_c - 1)}}{\log{(d_v - 1)}}}{1 - \frac{\log{2}}{\log{(d_v - 1)}}} \overset{d_c > d_v > 3}{\geq} 2.
	\end{align}
	It follows that if the transmit power is kept fixed even as $P_e \to 0$, the total power 
	diverges as $P_{\mathrm{total,bdd\;P_T}}=\Omega \left(\log^{2.48}\frac{1}{P_e} \right)$.
	Moving to the unbounded case, substituting~\eqref{eq:ratio} back into~\eqref{eq:galbtotal}, we obtain:
	\begin{eqnarray}
		\label{eq:galblowerboundfinal2}P_{\mathrm{total}} &=& \Omega\left(P_T + \left(\frac{\log\frac{1}{P_e}}{P_T}\right)^{2}\right).
	\end{eqnarray}
	Differentiating the expression inside the $\Omega \left( \cdot \right)$ on the RHS of~\eqref{eq:galblowerboundfinal2} 
	w.r.t. $P_T$ and setting to zero, the total power scales like:
	\begin{equation}
		\nonumber P_{\mathrm{total,min}}=\Omega\left(\log^\frac{2}{3} \frac{1}{P_e}\right).
	\end{equation}
	This lower bound tightens when $d_v d_c \geq 4(d_v + d_c)$. Using Theorem~\ref{thm:asymlbwire} for this case,
	\begin{eqnarray}
		\nonumber P_{\mathrm{total}} &=& \Omega\left(P_T + \left(\frac{\log\frac{1}{P_e}}{P_T}\right)^{\log 32}\right)\\
		\nonumber P_{\mathrm{total,min}} &\overset{\frac{\log 32}{\log 32 + 1} > \frac{31}{40}}{=}& \Omega\left(\log^\frac{31}{40} \frac{1}{P_e}\right).
	\end{eqnarray}
	Moving to the upper bound, via Theorem~\ref{thm:asymlbwire}, we find 
	that the exponent of $\frac{\log\frac{1}{P_e}}{P_T}$ in~\eqref{eq:galbtotalup} is
	\begin{equation}
		\label{eq:uppergammagallb} 2 \gamma \leq \frac{6\log (2 d_v d_c + 1)}{\log (d_v -1) - \log 2}.
	\end{equation}
	Then substituting~\eqref{eq:uppergammagallb} into~\eqref{eq:galbtotalup}, 
	we get the bound
	\begin{eqnarray}
		\label{eq:galbupperboundfinal}P_{\mathrm{total}} &=& \mathcal{O}\left(P_T + \left(\frac{\log\frac{1}{P_e}}{P_T}\right)^{\frac{6\log (2 d_v d_c + 1)}{\log (d_v - 1) - \log 2}}\right).
	\end{eqnarray}
	Differentiating the expressions inside the $\mathcal{O} \left( \cdot \right)$ of~\eqref{eq:galbupperboundfinal} 
	w.r.t. $P_T$ and setting to zero, we obtain the upper bound.
	\end{proofof}
	
	\section{CAD flow details}
	\label{app:cad}
	\allowdisplaybreaks
	The decoding implementation models are constructed in a hierarchical manner. First, behavioral 
	verilog descriptions of variable and check nodes are mapped to standard cells 
	using logic synthesis\footnote{The delay, power, area, and structure of 
	synthesized logic depend on the constraints and mapping effort given as inputs to the synthesis tool. To 
	allow for a fair comparison between codes of different degrees, we only specify constraints for minimum delay 
	and minimum power and use the highest possible mapping effort for each node.} and are then 
	placed-and-routed using a physical design tool. The physical area of the individual circuits is obtained.
	Post-layout simulation is then performed, using extracted RC parasitics and typical corners 
	for the Synopsys $32/28$nm HVT CMOS process at a supply voltage of $0.78$V. 
 	The critical-path delays of the variable-nodes $T_{\mathrm{VN}} (a, d_v)$, and check-nodes 
	$T_{\mathrm{CN}} (a, d_c)$ for each decoding algorithm are obtained using post-layout
	static timing analysis with the parasitics included. Post-layout power analysis is performed
	to obtain the average power consumption of variable-nodes $P_{\mathrm{VN}} (a, d_v)$ and check-nodes
	$P_{\mathrm{CN}} (a, d_c)$ using a ``virtual clock'' of period $T_{\mathrm{VN}} (a, d_v)$ or $T_{\mathrm{CN}} (a, d_c)$, 
	respectively, over a large number of uniformly random input patterns. In practice however, 
	the amount of switching activity at the decoder depends on the number of errors in the received 
	sequence over the channel, and it thereby depends on the parameters of the channel and communication 
	system. For example, when the transmit power is large and/or the path-loss and noise are small, the expected 
	number of errors in the received sequence is small and the switching activity caused by bit-flips may 
	be much smaller than these simulations indicate. Nevertheless, we assume (with slight overestimation) that
	the averaged power numbers hold for the various check-nodes and variable-nodes.
	
	\section{Circuit model for critical-path delay}
	\label{app:gatedelaymodels}
	It is assumed that all decoders operate at the minimum clock period 
	$T_{\mathrm{CLK}}(a, g, d_v, d_c)$ for which timing would be met at the $0.78$V supply voltage. 
	This minimum allowable clock period that meets timing in flip-flop based synchronous circuits 
	is bounded by the setup time constraint~\cite{JanBook} for each flip-flop. The setup time is the minimum time 
	it takes the incoming data to a flip-flops to propagate through the input stages of the flip-flop. The critical path 
	for a full decoding iteration consists of a CLK-Q delay of a message passing flip-flop inside a variable-node, 
	then an interconnect delay, then a check-node delay $T_{\mathrm{CN}} (a, d_c)$, then another interconnect 
	delay, and finally a variable-node delay $T_{\mathrm{VN}} (a, d_v)$. In these models, the setup and the 
	CLK-Q delay are accounted for in $T_{\mathrm{VN}} (a, d_v)$.

	Interconnect delay is assumed to be linearly proportional to the resistance and 
	capacitance of the interconnect, which depend on the length and width of interconnects. Estimating 
	the length of interconnects requires an estimate of the decoder's physical dimensions. 
	The total area of the decoder, $A_{\mathrm{Decoder}}$, is estimated as a sum of check-node and variable-node 
	areas, where the nodes are assumed to be placed in a square arrangement. Best-case and worst-case estimates 
	for the average interconnect length $l_{\mathrm{wire}}(a, g, d_v, d_c)$ are obtained by the following equations~\cite{rentrule1}
	\begin{numcases}{l_{\mathrm{wire}}(a, g, d_v, d_c)=}
	A_{\mathrm{Decoder}}^{0.25} & in best case. \label{eq:rentruleparallel} \\
	\frac{\sqrt{A_{\mathrm{Decoder}}}}{3} & in worst case. \label{eq:rentrulerandom}
	\end{numcases}
	
	Rigorous empirical and theoretical justification for the above estimates is 
	provided in~\cite{rentrule1} where it is shown that~\eqref{eq:rentruleparallel} is a good approximation for 
	highly-parallel logic and~\eqref{eq:rentrulerandom} is the average value for randomly-placed logic
	on a square array. Since the logic functions computed by the check-nodes and variable-nodes for the 
	decoding algorithms considered in this paper are intrinsically parallel and we also assume the decoders are 
	implemented in a fully-parallel manner, we used~\eqref{eq:rentruleparallel} for the results shown in this paper.
	However, we note that~\eqref{eq:rentrulerandom} could be a better approximation, depending on
	the code construction used.
	
	Routing for decoders is assumed to use minimum-width wires on the lower $7$ metal layers of the $9$-layer CMOS 
	process\footnote{The top two metal layers are often used to construct a global power grid for 
	an entire chip.}. The average minimum width $(w_{\mathrm{avg}})$, sheet resistance 
	$(R_{\mathrm{sq}})$, and capacitance per-unit-length $(C_{\mathrm{unit}})$ \footnote{Including 
	parallel-plate and fringing components~\cite{JanBook}.} for these metal layers 
	are calculated using design rule information~\cite{SynopsysLib} and are assumed as constants. 
	Interconnect delay is then estimated assuming a distributed Elmore model~\cite{JanBook}:
	\begin{eqnarray}
		R_{\mathrm{wire}} (a, g, d_v, d_c) &=& R_{\mathrm{sq}} \times \frac{l_{\mathrm{wire}}(a, g, d_v, d_c)}{w_{\mathrm{avg}}}\\
		C_{\mathrm{wire}} (a, g, d_v, d_c) &=& C_{\mathrm{unit}} \times l_{\mathrm{wire}}(a, g, d_v, d_c)\\
		T_{\mathrm{wire}} (a, g, d_v, d_c) &=& \frac{R_{\mathrm{sq}} C_{\mathrm{unit}} l^{2}_{\mathrm{wire}}(a, g, d_v, d_c)}{2 w_{\mathrm{avg}}}.
	\end{eqnarray} 
	
	\section{Circuit model for computation power}
	\label{app:nodepowersim}
	\allowdisplaybreaks
	The power consumption of a logic gate consists of both dynamic power 
	(which is proportional to the activity-factor at the input of the gate and the clock-frequency), 
	and static power (which has no dependence on the activity-factor or the clock frequency)~\cite{JanBook}. 
	In post-layout simulation, the static power consumption of variable-nodes and 
	check-nodes at $0.78$V supply in a high threshold-voltage process is observed 
	to be less than $1\%$ of the total power in check-nodes and variable-nodes.
	Therefore, with little loss in accuracy, we treat the total power consumption of check-nodes 
	and variable-nodes as dynamic power when considering the effect of 
	clock-frequency scaling. Therefore, the power consumed after clock-frequency scaling 
	in variable-nodes is $P_{\mathrm{VN}}(a, d_v) \times \frac{T_{\mathrm{VN}} (a, d_v)}{T_{\mathrm{CLK}}(a, g, d_v, d_c)}$
	and in check-nodes it is $P_{\mathrm{CN}}(a, d_c) \times \frac{T_{\mathrm{VN}} (a, d_v)}{T_{\mathrm{CLK}}(a, g, d_v, d_c)}$.
	
	\section{Circuit model for interconnect power}
	\label{app:wirepowersim}
	Using the interconnect capacitance estimate $C_{\mathrm{wire}}(a, g, d_v, d_c)$ and clock 
	period $T_{\mathrm{CLK}} (a, g, d_v, d_c)$ from Appendix~\ref{app:gatedelaymodels}, and 
	assuming an activity factor of $\frac{1}{2}$, the 	power consumed by a single message-passing 
	interconnect $(P_{\mathrm{wire}}(a, g, d_v, d_c))$ in the decoder is modeled using the formula for the dynamic power consumed in 
	interconnects~\cite{JanBook}:
	\begin{equation}
		P_{\mathrm{wire}} (a, g, d_v, d_c) = \frac{C_{\mathrm{wire}} (a, g, d_v, d_c) \times (0.78 V)^2}{2 T_{\mathrm{CLK}} (a, g, d_v, d_c)}.
	\end{equation} 

	\bibliographystyle{IEEEtran}
	{\footnotesize \bibliography{IEEEabrv,sources}}

	\begin{IEEEbiography}[{\includegraphics[width=1in,height=1.25in,clip,keepaspectratio]{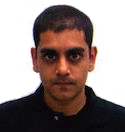}}]{Karthik Ganesan}
	received the B.S. degree in EECS and the B.A. degree in Statistics from the 
	University of California at Berkeley in 2013, and the M.S. degree in EE from 
	Stanford University in 2015, where he is currently pursuing the Ph.D. degree. He is 
	interested in some aspects of coding theory, applied probability and ergodic theory, 
	and their uses in fault-tolerant computing, low-power system design, and
	emerging neuroscience applications.
	\end{IEEEbiography}

	\vspace*{-2\baselineskip}
	\begin{IEEEbiography}[{\includegraphics[width=1in,clip]{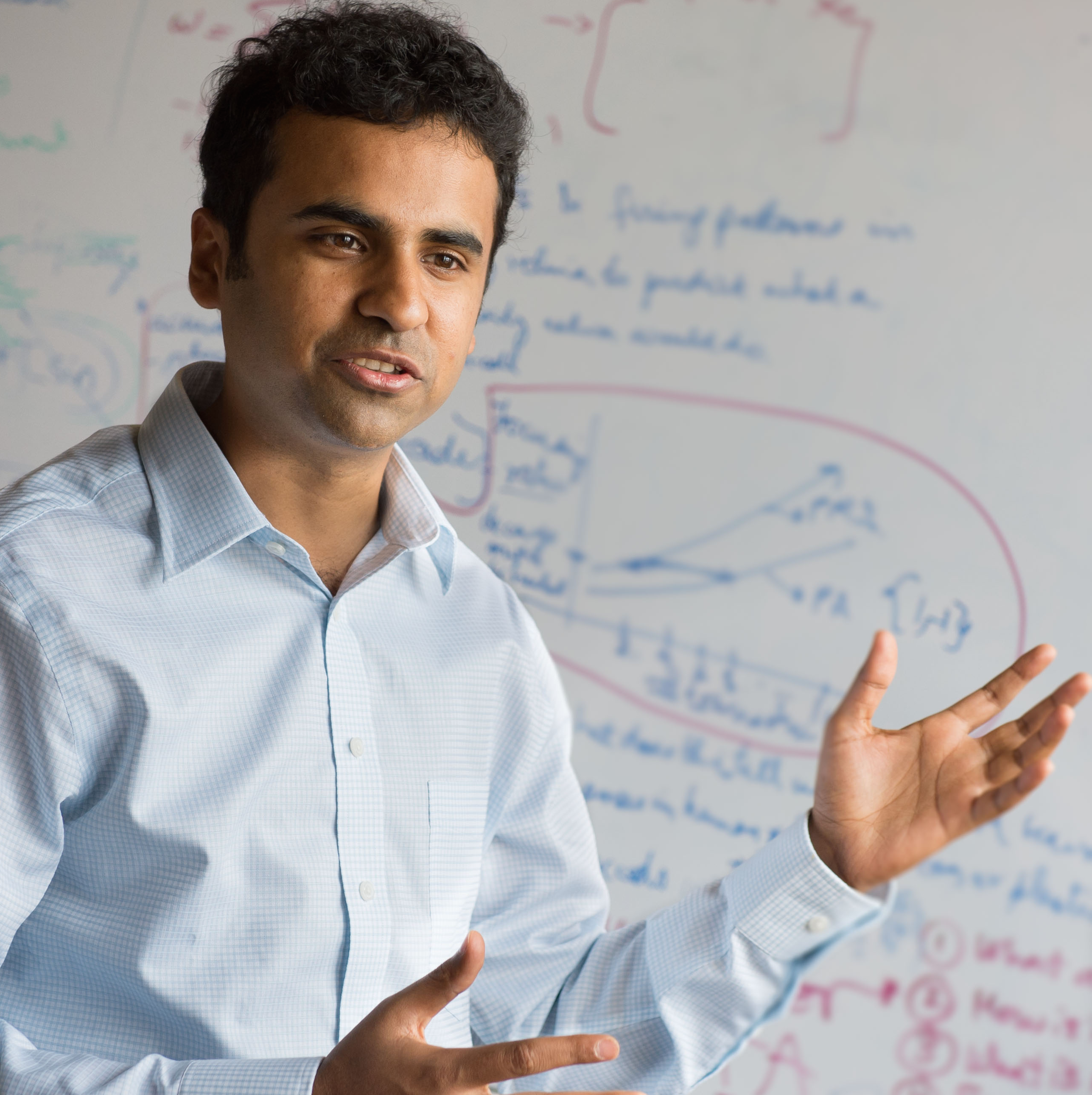}}]{Pulkit Grover}
	is an assistant professor in Electrical and Computer Engineering at 
	Carnegie Mellon University (since 2013), working on information theory, 
	circuit design, and biomedical engineering. His focus is on developing a new 
	theory of information for low-energy communication, sensing, and computing by 
	incorporating novel circuit/processing-energy models to add to classical communication 
	or sensing energy models. A common theme in his work is observing when 
	optimal designs depart radically from classical theoretical intuition. To apply these ideas to a 
	variety of problems including wearables, IoT, and novel biomedical systems, 
	his lab works extensively with engineers, neuroscientists, and doctors. 
	
	He is a recipient of the 2010 best student paper award at the 
	IEEE Conference on Decision and Control; a 2010 best student paper 
	finalist at the IEEE International Symposium on Information Theory; the 
	2011 Eli Jury Dissertation Award from UC Berkeley; the 2012 
	Leonard G. Abraham award from the IEEE Communications Society; a 2014 
	best paper award at the International Symposium on Integrated Circuits; a 
	2014 NSF CAREER award; and a 2015 Google Research Award.
	\end{IEEEbiography}

	\vspace*{-2\baselineskip}
	\begin{IEEEbiography}[{\includegraphics[width=1in,height=1.25in,clip,keepaspectratio]{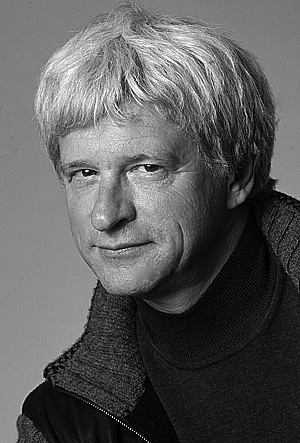}}]{Jan Rabaey}
	is the Donald O. Pederson Distinguished Professor in the Electrical Engineering and 
	Computer Science Department, University of California at Berkeley. He is currently 
	the Scientific Co-director of the Berkeley Wireless Research Center (BWRC), the 
	director of the Berkeley Ubiquitous SwarmLab, and the Director of the FCRP Multiscale 
	Systems Research Center (MuSyC). His research interests include the conception and 
	implementation of next-generation integrated wireless systems.
  
	Dr. Rabaey is the recipient of a wide range of awards, among which are
	the 2008 IEEE Circuits and Systems Society Mac Van Valkenburg Award and
	the 2009 European Design Automation Association (EDAA) Lifetime Achievement
	award. In 2010, he was awarded the prestigious Semiconductor Industry
	Association (SIA) University Researcher Award. He is an IEEE Fellow and a member of the Royal 
	Flemish Academy of Sciences and Arts of Belgium. He received his Ph.D. degree 
	in applied sciences from the Katholieke Universiteit Leuven, Leuven, Belgium.
	\end{IEEEbiography}

	\vspace*{-2\baselineskip}
	\begin{IEEEbiography}[{\includegraphics[width=1in,height=1.25in,clip,keepaspectratio]{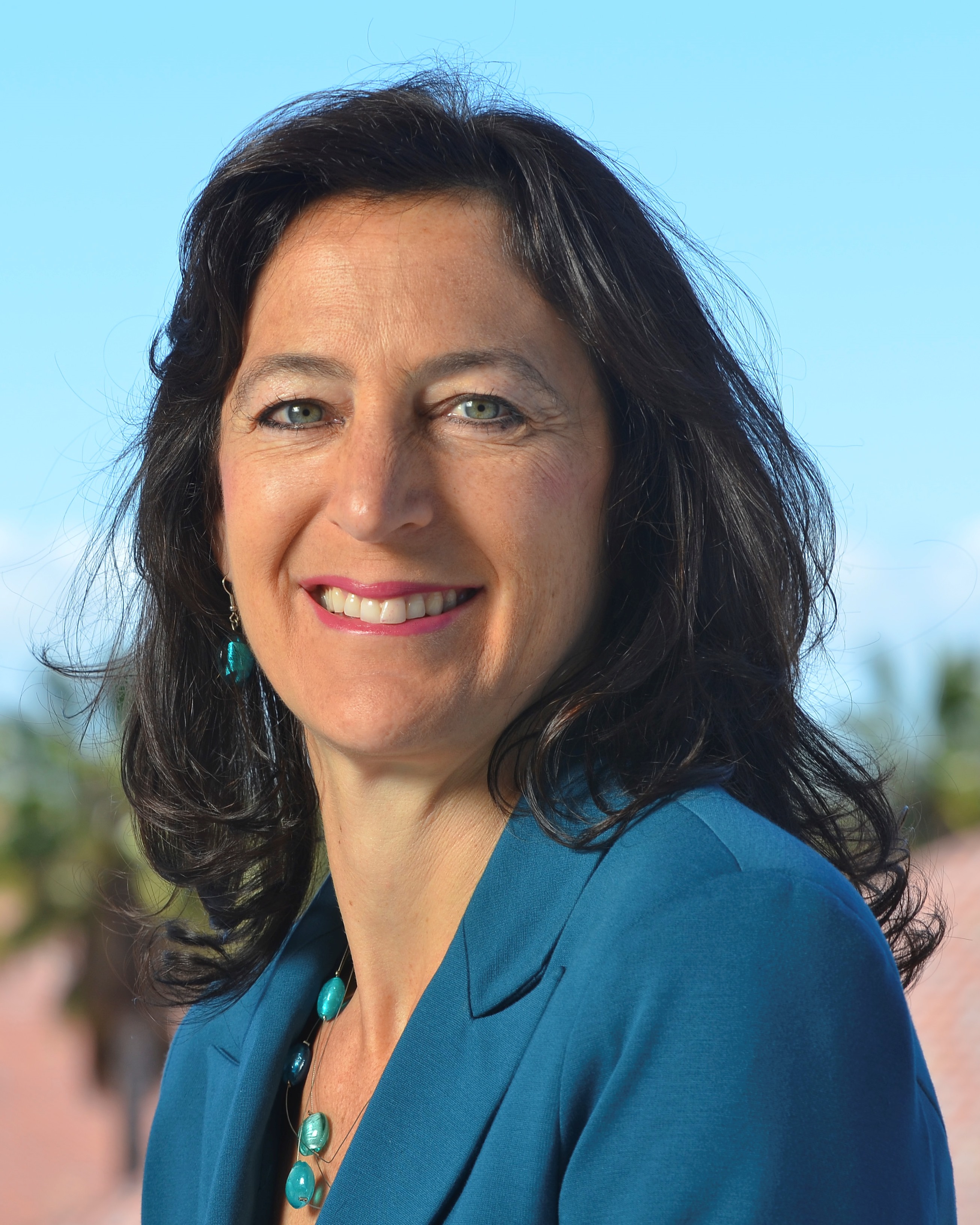}}]{Andrea Goldsmith}
	is the Stephen Harris professor in the School of Engineering and a professor of 
	Electrical Engineering at Stanford University. She was previously on the faculty of 
	Electrical Engineering at Caltech. Her research interests are in information theory and 
	communication theory, and their application to wireless communications and related fields. 
	She co-founded and served as Chief Scientist of Wildfire.Exchange, and previously co-founded 
	and served as CTO of Quantenna Communications, Inc. She has also held industry positions 
	at Maxim Technologies, Memorylink Corporation, and AT\&T Bell Laboratories. Dr. Goldsmith is a 
	Fellow of the IEEE and of Stanford, and has received several awards for her work, 
	including the IEEE ComSoc Edwin H. Armstrong Achievement Award as well as Technical Achievement 
	Awards in Communications Theory and in Wireless Communications, the National Academy of Engineering 
	Gilbreth Lecture Award, the IEEE ComSoc and Information Theory Society Joint Paper Award, the 
	IEEE ComSoc Best Tutorial Paper Award, the Alfred P. Sloan Fellowship, the WICE Technical Achievement 
	Award, and the Silicon Valley/San Jose Business Journal's Women of Influence Award. She is author of the book 
	``Wireless Communications'' and co-author of the books ``MIMO Wireless Communications'' and ``Principles of 
	Cognitive Radio,'' all published by Cambridge University Press, as well as an inventor on 28 patents. She received the 
	B.S., M.S. and Ph.D. degrees in Electrical Engineering from U.C. Berkeley.

	Dr. Goldsmith has served on the Steering Committee for the IEEE Transactions on Wireless Communications 
	and as editor for the IEEE Transactions on Information Theory, the Journal on Foundations and Trends in 
	Communications and Information Theory and in Networks, the IEEE Transactions on Communications, and the 
	IEEE Wireless Communications Magazine. She participates actively in committees and conference organization 
	for the IEEE Information Theory and Communications Societies and has served on the Board of Governors for both 
	societies. She has also been a Distinguished Lecturer for both societies, served as President of the IEEE Information 
	Theory Society in 2009, founded and chaired the student committee of the IEEE Information Theory society, and 
	chaired the Emerging Technology Committee of the IEEE Communications Society. At Stanford she received the 
	inaugural University Postdoc Mentoring Award, served as Chair of Stanford's Faculty Senate in 2009 and currently 
	serves on its Faculty Senate, Budget Group, and Task Force on Women and Leadership.
	\end{IEEEbiography}
	
\end{document}